\RequirePackage{fix-cm}
\documentclass[smallextended]{svjour3}
\smartqed
\usepackage{graphicx}

\usepackage{amsmath}
\usepackage{amssymb}
\usepackage{braket}
\usepackage{color}

\newtheorem{thm}{Theorem}

\begin{document}

\title{Some limit laws for quantum walks with applications to a version of the Parrondo paradox}
\subtitle{}

\titlerunning{Some limit laws for QWs with applications to a version of the Parrondo paradox}

\author{Takuya Machida \and F. Alberto Gr{\"u}nbaum}

\authorrunning{T.~Machida \and F. A. Gr{\"u}nbaum}

\institute{%
T.~Machida \at
              College of Industrial Technology, Nihon University, Narashino, Chiba 275-8576, Japan\\
              \email{machida.takuya@nihon-u.ac.jp}\\
F. Alberto Gr{\"u}nbaum \at
              Department of Mathematics, University of California, Berkeley, CA, 94720, USA\\
              \email{grunbaum@math.berkeley.edu}
}

\date{}

\maketitle

\begin{abstract}
A quantum walker moves on the integers with four extra degrees of freedom, performing a coin-shift operation to alter its internal state and position at discrete units of time. 
The time evolution is described by a unitary process.
We focus on finding the limit probability law for the position of the walker and study it by means of Fourier analysis.
The quantum walker exhibits  both localization and a ballistic behavior.
Our two results are given as limit theorems for a 2-period time-dependent walk and they describe the location of the walker after it has repeated the unitary process a large number of times.
The theorems  give an analytical tool to study some of the Parrondo type behavior in a quantum game which was studied by J. Rajendran and C. Benjamin by means of very nice numerical simulations~\cite{RajendranBenjamin2018}.
With our analytical tools at hand we can easily explore the ``phase space'' of parameters of one of the games, similar to the winning game in their papers. We include numerical evidence that our two games, similar to theirs, exhibit a Parrondo type paradox.

\keywords{Quantum walk \and Limit theorem \and Parrondo paradox}
\end{abstract}

\section{Introduction}
Quantum walks, introduced in~\cite{ADZ}, can be considered as a counterpart of random walks and have been studied in mathematics since around 2000.
With spin orientations, also known as coin states, the quantum walkers move in a discrete space in superposition.
The position of a quantum walker has a limiting distribution which is very far  from those of classical random walks.
In contrast to more familiar classical random walks, some of the quantum walkers localize.
Quantum walks can be useful to build quantum search algorithms which are expected to be working in quantum computers~\cite{VK,Venegas-Andraca2008,Venegas-Andraca2012}.

We study a quantum walk with four spin orientations and analyze the probability distribution for its position.
The quantum walker is defined on the line $\mathbb{Z}=\left\{0,\pm 1,\pm2,\ldots\right\}$ and its time evolution is given by iterating a unitary transformation.
We establish its probability distribution as two limit theorems resulting in localization.
One of them is a weak limit result and the other one gives convergence in distribution.
Both theorems give us  an indication of the behavior of the walker for large times.

Our limit theorems give part of the tools needed to validate a Parrondo type paradox in a quantum game.
The quantum game was introduced in Rajendran and Benjamin~\cite{RajendranBenjamin2018}, and they found by numerical simulations a result similar to one that we will study after giving the proofs of our theorems.
The game defined by Rajendran and Benjamin is a quantum walk with four spin orientations and the walker shifts to the left or right, or stays at the same position depending on the value of its spin state.
The quantum walk is a time-dependent walk and they focused on a numerical study on a 2-period time-dependent quantum walk and compared it to a 1-period quantum walk. Our results deal with a 2-period time-dependent walk.
To keep the length of this paper reasonable, we postpone the derivation of other analytical results that are needed to give a complete validation of the numerical results in~\cite{RajendranBenjamin2018}.

Studies previous to this one include a limit distribution of a 2-period time-dependent quantum walk with two spin orientations on the line revealing that the walker delocalizes in distribution~\cite{MachidaKonno2010}, as well as two limit theorems for a quantum walk with four spin orientations on the line showing  that the quantum walk could localize~\cite{KonnoMachida2010}.

\bigskip

The  paper starts with the definition of the quantum walk on the line. 
We get two limit theorems and Sec.~\ref{sec:limit_measure} is devoted to establishing a limit measure for the quantum walk.
The other limit theorem is given in Sec.~\ref{sec:convergence_in_distribution}.
Both of them reveal the important fact that the quantum walk can localize in distribution.  
Section~\ref{sec:application} displays analytical results that are then used to display regions of parameter space where our 2-period time-dependent quantum walk is a winning or a losing game. In section $6$ we give numerical evidence of a Parrondo type
paradox for choice of coins that is simpler than the one in
\cite{RajendranBenjamin2018}.
The paper closes with a summary.

\section{Definition}
\label{sec:definition}
\begin{figure}[h]
\begin{center}
 \begin{minipage}{50mm}
  \begin{center}
  \includegraphics[scale=0.4]{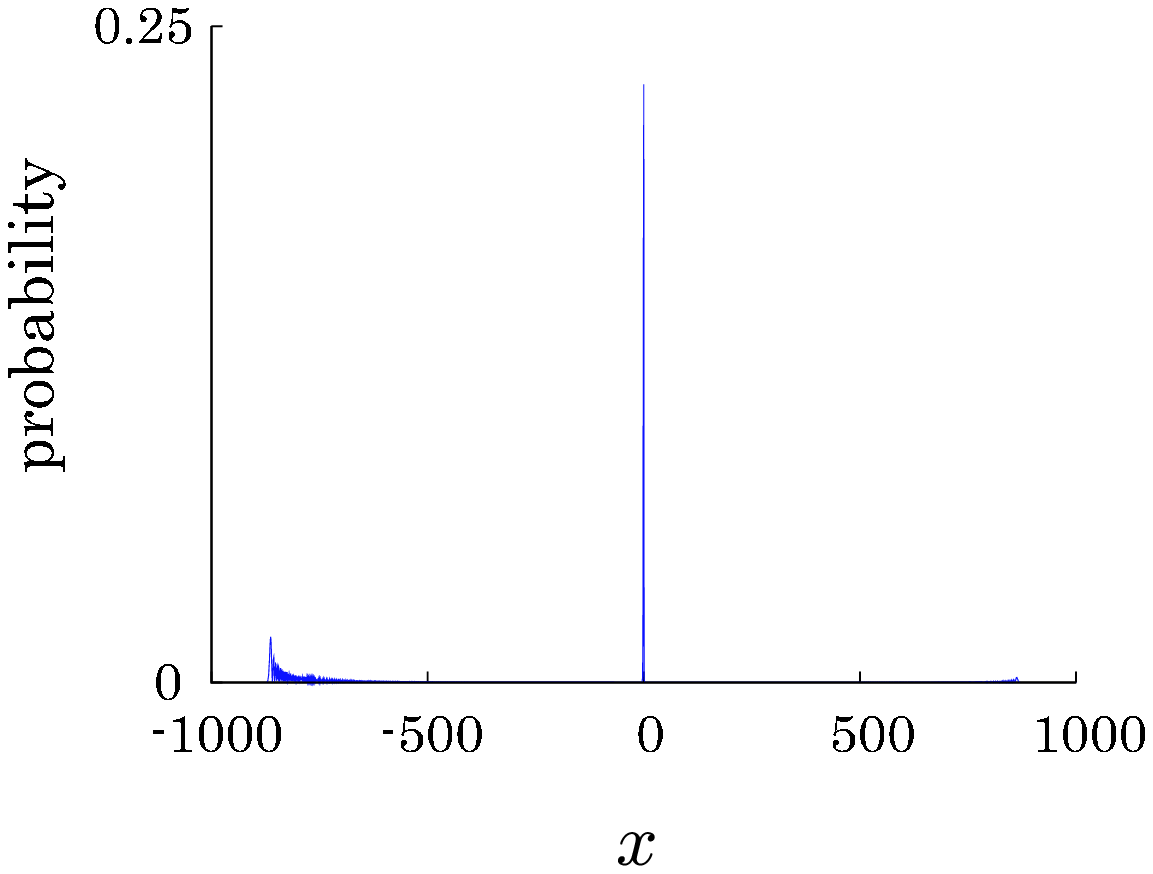}\\[2mm]
  (a) $\theta_1=\pi/6, \theta_2=\pi/6$
  \end{center}
 \end{minipage}
 \begin{minipage}{50mm}
  \begin{center}
 \includegraphics[scale=0.4]{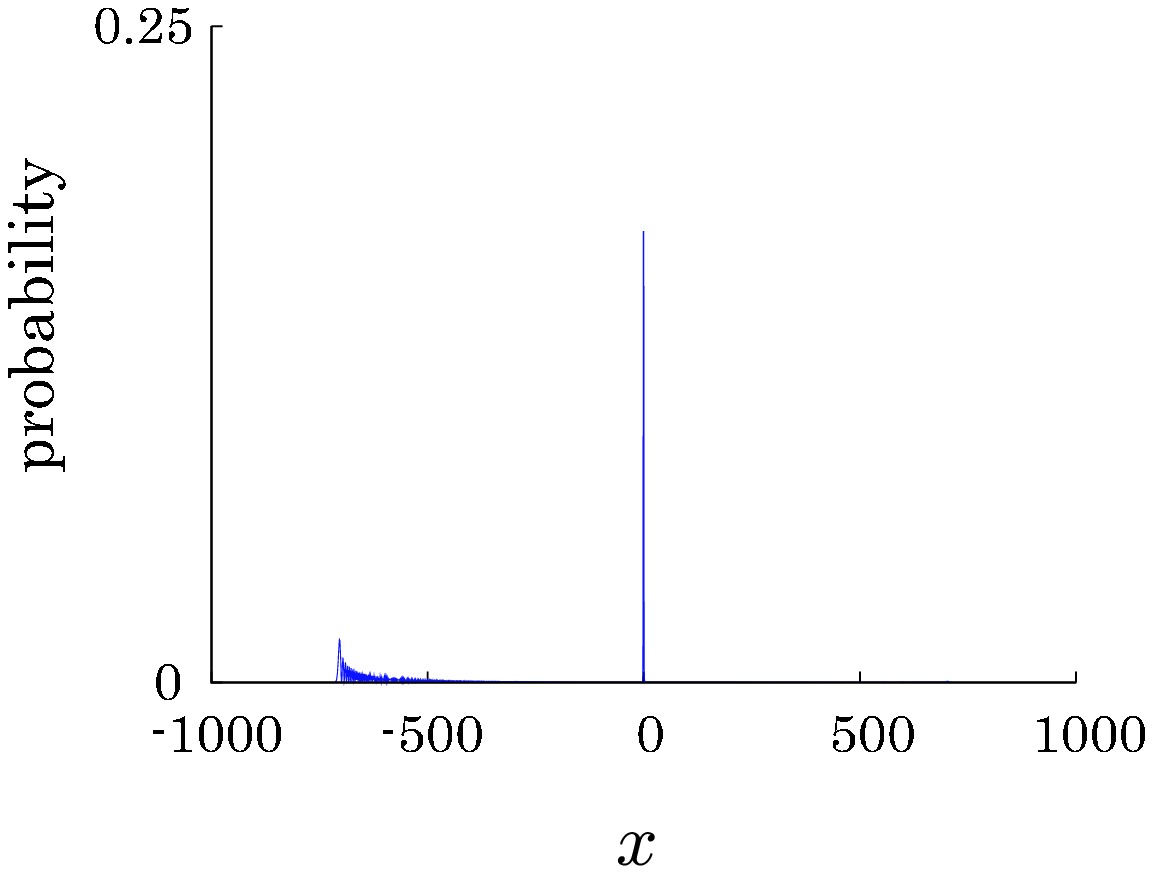}\\[2mm]
  (b) $\theta_1=\pi/4, \theta_2=\pi/4$
  \end{center}
 \end{minipage}
 \vspace{5mm}

 \begin{minipage}{50mm}
  \begin{center}
  \includegraphics[scale=0.4]{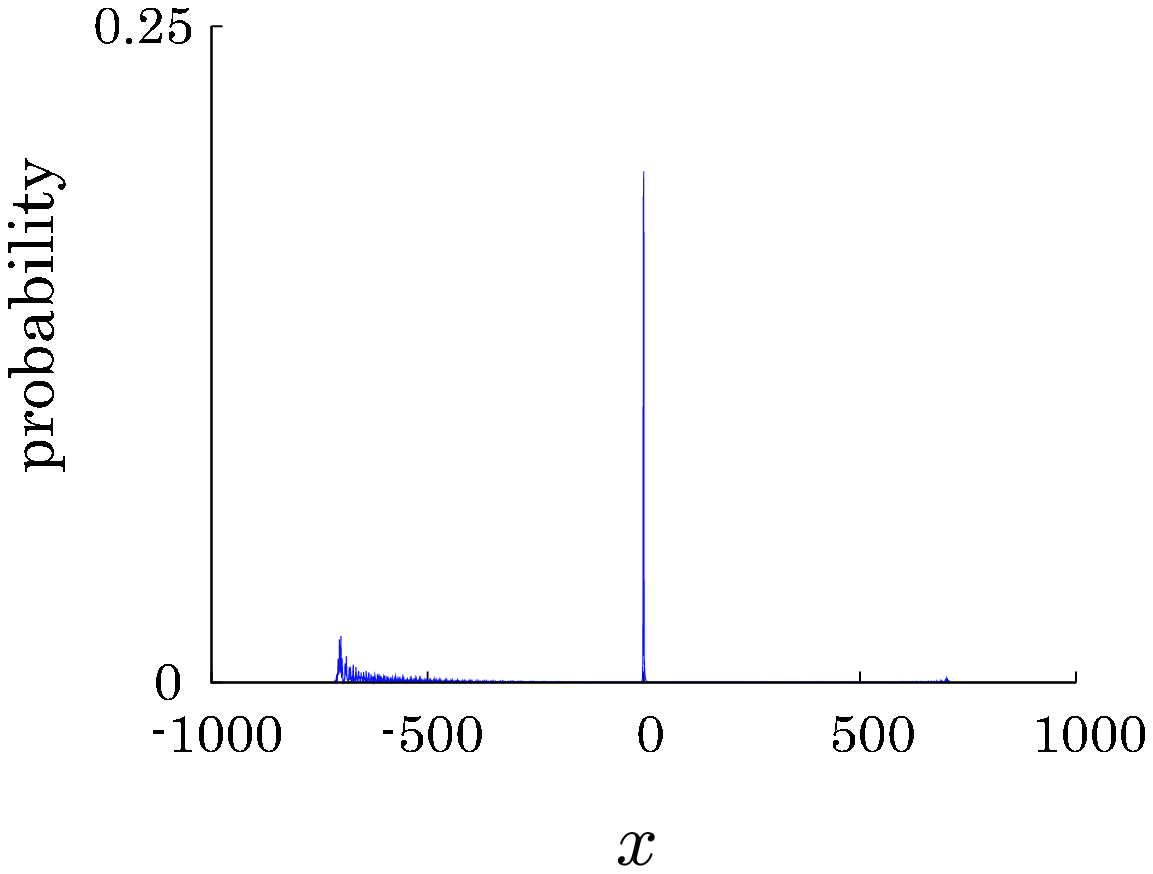}\\[2mm]
  (c) $\theta_1=\pi/6, \theta_2=\pi/4$
  \end{center}
 \end{minipage}
 \begin{minipage}{50mm}
  \begin{center}
 \includegraphics[scale=0.4]{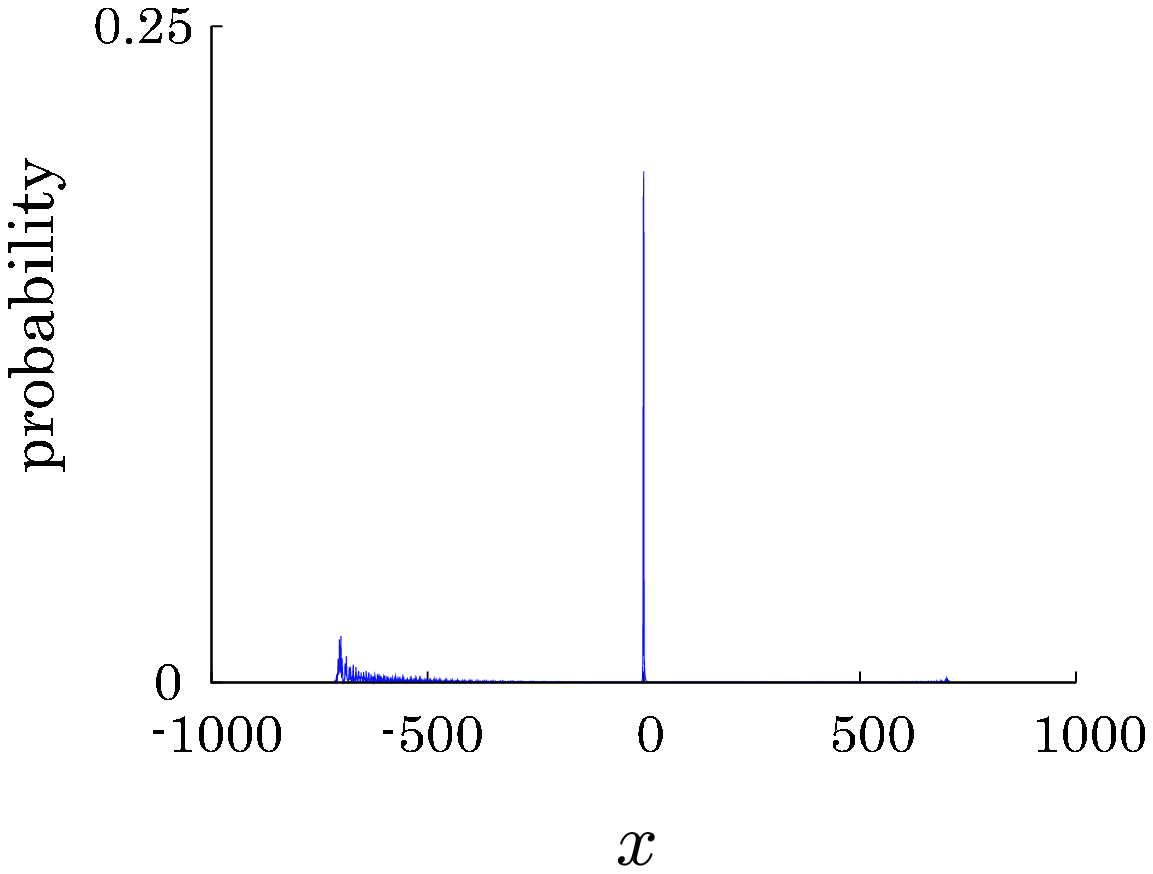}\\[2mm]
  (d) $\theta_1=\pi/4, \theta_2=\pi/6$
  \end{center}
 \end{minipage}
 \vspace{5mm}
 \caption{The quantum walker partly localizes in probability distribution at time $500$, $\mathbb{P}(X_{500}=x)$ with $q_0=q_1=q_2=q_3=1/2$.}
 \label{fig:fig1}
\end{center}
\end{figure}

The quantum walker has four coin states, represented by $\ket{00}, \ket{01}, \ket{10}$, and $\ket{11}$, and these states are manipulated with a coin-flip operation and a position-shift operation.
The quantum walk is described in a tensor Hilbert space $\mathcal{H}_p\otimes\mathcal{H}_c$,
\begin{equation}
 \ket{\Psi_t}=\sum_{x\in\mathbb{Z}} \ket{x}\otimes \ket{\psi_t(x)}\,\,\in\mathcal{H}_p\otimes\mathcal{H}_c,
\end{equation}
where the Hilbert space $\mathcal{H}_p$ is spanned by the orthogonal basis $\left\{\ket{x} : x\in\mathbb{Z}\right\}$ and the Hilbert space $\mathcal{H}_c$ is spanned by the orthogonal basis $\left\{\ket{00}, \ket{01}, \ket{10}, \ket{11}\right\}$.
With two coin-flip operations and a position shift-operation which are all unitary operations, the state at time $t$ evolves into the state at time $t+1$ as follows, 
\begin{equation}
 \ket{\Psi_{t+1}}=SYSX\ket{\Psi_t},\label{eq:time_evolution}
\end{equation}
where the coin-flip operations $X$ and $Y$, and the position-shift operation $S$ are given explicitly by
\begin{align}
 X=&\sum_{x\in\mathbb{Z}}\ket{x}\bra{x}\otimes (U_2\otimes U_1),\label{eq:operation_X}\\
 Y=&\sum_{x\in\mathbb{Z}}\ket{x}\bra{x}\otimes (U_1\otimes U_2),\label{eq:operation_Y}\\
 U_1=&c_1\ket{0}\bra{0}+s_1\ket{0}\bra{1}+s_1\ket{1}\bra{0}-c_1\ket{1}\bra{1},\\
 U_2=&c_2\ket{0}\bra{0}+s_2\ket{0}\bra{1}+s_2\ket{1}\bra{0}-c_2\ket{1}\bra{1},\\
 S=&\sum_{x\in\mathbb{Z}}\ket{x-1}\bra{x}\otimes\ket{00}\bra{00}+\ket{x}\bra{x}\otimes\ket{01}\bra{01}\nonumber\\
 &+\ket{x}\bra{x}\otimes\ket{10}\bra{10}+\ket{x+1}\bra{x}\otimes\ket{11}\bra{11},
\end{align}
with $c_1=\cos\theta_1, s_1=\sin\theta_1, c_2=\cos\theta_2, s_2=\sin\theta_2\, (\theta_1,\theta_2\in [0,2\pi))$.
We assume that $\theta_1,\theta_2\notin\left\{0,\pi/2,\pi,3\pi/2\right\}$.
Note that $\ket{j_1}\otimes\ket{j_2}=\ket{j_1j_2}\,(j_1,j_2\in\left\{0,1\right\})$.
Equation~\eqref{eq:time_evolution} can be also considered to be a two-step quantum walk, and the operations $SX$ and $SY$ are 2-periodically repeated,
\begin{equation}
 \ket{\Psi_t}=SYSX\ket{\Psi_{t-1}}=SYSXSYSX\ket{\Psi_{t-2}}=\cdots=(SYSX)^t\ket{\Psi_0}.
\end{equation}
The quantum walker is assumed to start at a localized initial state
\begin{equation}
 \ket{\Psi_0}=\ket{0}\otimes\Bigl(q_0\ket{00}+q_1\ket{01}+q_2\ket{10}+q_3\ket{11}\Bigr),\label{eq:initial_state}
\end{equation}
with complex numbers $q_0, q_1, q_2, q_3$ such that $|q_0|^2+|q_1|^2+|q_2|^2+|q_3|^2=1$.
The quantum walker is observed at position $x$ at time $t$ with probability distribution for its position given by
\begin{equation}
 \mathbb{P}(X_t=x)=\bra{\Psi_t}\left(\ket{x}\bra{x}\otimes\sum_{j_1=0}^1\sum_{j_2=0}^1\ket{j_1j_2}\bra{j_1j_2}\right)\ket{\Psi_t}.
\end{equation}
Figure~\ref{fig:fig1} gives some instances of this probability distribution at time $500$ by numerical experiments on which we find the quantum walker localizing around the origin.
The quantum walker has four spin orientations, but similar kind of probability distributions arise even for a quantum walk with three spin orientations as well~\cite{Machida2015}.

\section{Limit measure}
\label{sec:limit_measure}
Since the numerical experiments suggest localization of the quantum walk, we expect an interesting limit of the probability distribution $\mathbb{P}(X_t=x)$ as $t\to\infty$.
It is indeed possible to compute the limit.

\begin{thm}
\label{th:limit_theorem_1}
Given the localized initial state in Eq.~\eqref{eq:initial_state}, the probability $\mathbb{P}(X_t=x)$ of finding the particle at location $x$ at time $t$ converges to the following expression as $t\to\infty$, 
\begin{align}
 \lim_{t\to\infty}\mathbb{P}(X_t=x)=&\Bigl|\xi_1(x ; q_0, q_1, q_2, q_3)\Bigr|^2+\Bigl|\xi_1(-x ; q_3, -q_2, -q_1, q_0)\Bigr|^2\nonumber\\
 &+\Bigl|\xi_2(x ; q_0, q_1, q_2, q_3)\Bigr|^2+\Bigl|\xi_2(-x ; -q_3, q_2, q_1, -q_0)\Bigr|^2,\label{eq:limit_measure}
\end{align}
where $\xi_1,\xi_2$ stand for
\begin{align}
 \xi_1(x ; q_0, q_1, q_2, q_3)=&\eta_1(x)q_0-\eta_3(x)q_1-\eta_3(-x-1)q_2-\eta_1(x+1)q_3,\\
 \xi_2(x ; q_0, q_1, q_2, q_3)=&-\eta_3(-x)q_0+\left(\delta_{0x}-\eta_1(x)\right)q_1+\eta_2(-x)q_2\nonumber\\
 &+\eta_3(-x-1)q_3,
\end{align}
and $\eta_1,\eta_2, \eta_3$ are given by
\begin{align}
 \eta_1(x)=&-\frac{1}{4}\Biggl[(s_1-s_2)\biggl\{\nu_1^{|x|}I(s_1<s_2)-\nu_1^{-|x|}I(s_1>s_2)\biggr\}\nonumber\\
 &+(s_1+s_2)\biggl\{\nu_2^{|x|}I(s_1<-s_2)-\nu_2^{-|x|}I(s_1>-s_2)\biggr\}\Biggr],
\end{align}
\begin{align}
 \eta_2(x)=&\left\{%
 \begin{array}{l}
  \frac{1}{4}\Biggl[(s_1-s_2)\biggl\{\nu_2\nu_1^{x}I(s_1<s_2)-\nu_2^{-1}\nu_1^{-x}I(s_1>s_2)\biggr\}\\
  +(s_1+s_2)\biggl\{\nu_1\nu_2^{x}I(s_1<-s_2)-\nu_1^{-1}\nu_2^{-x}I(s_1>-s_2)\biggr\}\Biggr]\\
  \hspace{65mm}(x=0,1,2,\ldots),\\
  \frac{1}{4}\Biggl[(s_1-s_2)\biggl\{\nu_2^{-1}\nu_1^{-x}I(s_1<s_2)-\nu_2\nu_1^{x}I(s_1>s_2)\biggr\}\\
  +(s_1+s_2)\biggl\{\nu_1^{-1}\nu_2^{-x}I(s_1<-s_2)-\nu_1\nu_2^{x}I(s_1>-s_2)\biggr\}\Biggr]\\
  \hspace{65mm}(x=-1,-2,\ldots),
 \end{array}
 \right.
\end{align}
\begin{align}
 \eta_3(x)=&\left\{%
 \begin{array}{l}
  \frac{c_1}{4}\Biggl[(s_1-s_2)\biggl\{\frac{1}{1-s_1}\nu_1^{x}I(s_1<s_2)+\frac{1}{1+s_1}\nu_1^{-x}I(s_1>s_2)\biggr\}\\
  +(s_1+s_2)\biggl\{\frac{1}{1-s_1}\nu_2^{x}I(s_1<-s_2)+\frac{1}{1+s_1}\nu_2^{-x}I(s_1>-s_2)\biggr\}\Biggr]\\
  \hspace{65mm}(x=0,1,2,\ldots),\\
  -\frac{c_1}{4}\Biggl[(s_1-s_2)\biggl\{\frac{1}{1+s_1}\nu_1^{-x}I(s_1<s_2)+\frac{1}{1-s_1}\nu_1^{x}I(s_1>s_2)\biggr\}\\
  +(s_1+s_2)\biggl\{\frac{1}{1+s_1}\nu_2^{-x}I(s_1<-s_2)+\frac{1}{1-s_1}\nu_2^{x}I(s_1>-s_2)\biggr\}\Biggr]\\
  \hspace{65mm}(x=-1,-2,\ldots),
 \end{array}
 \right.
\end{align}
\end{thm}

Above we have used
\begin{equation}
 \nu_1=\frac{(1+s_1)(1-s_2)}{c_1c_2},\quad \nu_2=-\frac{(1+s_1)(1+s_2)}{c_1c_2},
\end{equation}
\begin{align}
 I(P)=&\left\{\begin{array}{ll}
	  1& (\mbox{if $P$ is true})\\
	       0& (\mbox{otherwise})
		\end{array}\right.,\\
 \delta_{0x}=&\left\{\begin{array}{ll}
	      1& (x=0)\\
		     0& (x=\pm 1, \pm 2,\ldots)
		    \end{array}\right..
\end{align}

\begin{figure}[h]
\begin{center}
 \begin{minipage}{50mm}
  \begin{center}
  \includegraphics[scale=0.4]{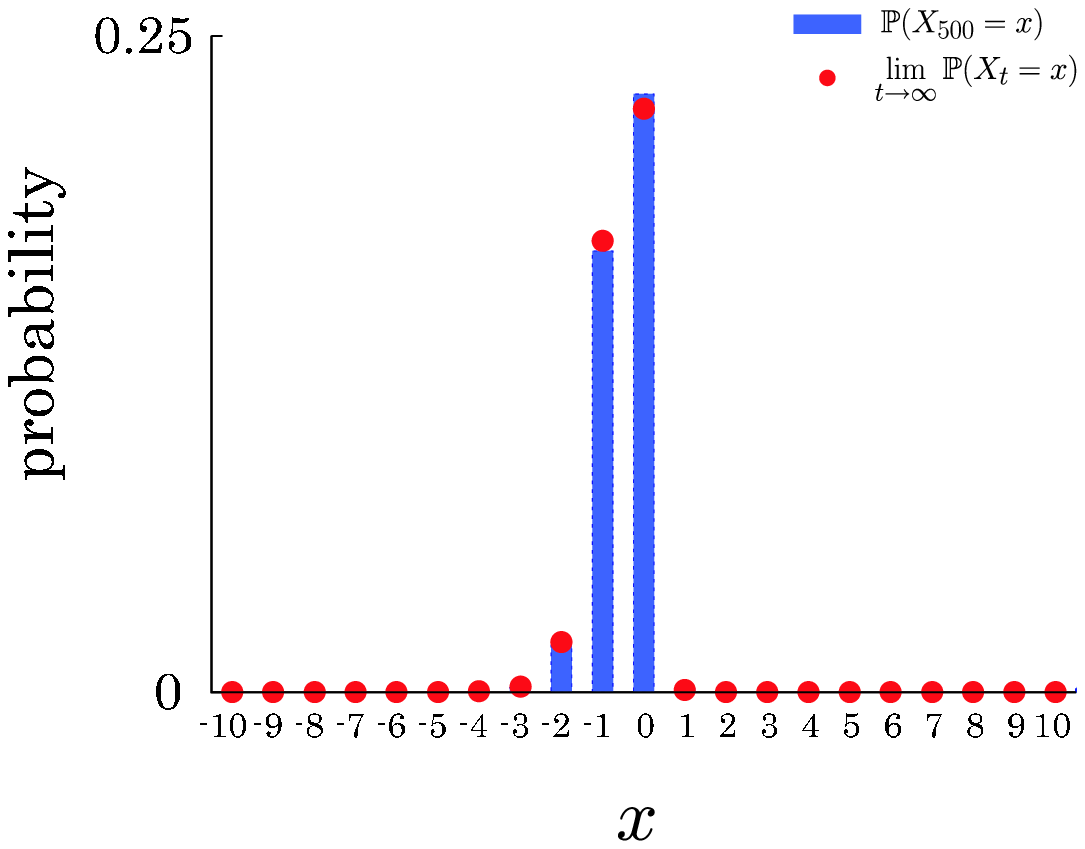}\\[2mm]
  (a) $\theta_1=\pi/6, \theta_2=\pi/6$
  \end{center}
 \end{minipage}
 \begin{minipage}{50mm}
  \begin{center}
 \includegraphics[scale=0.4]{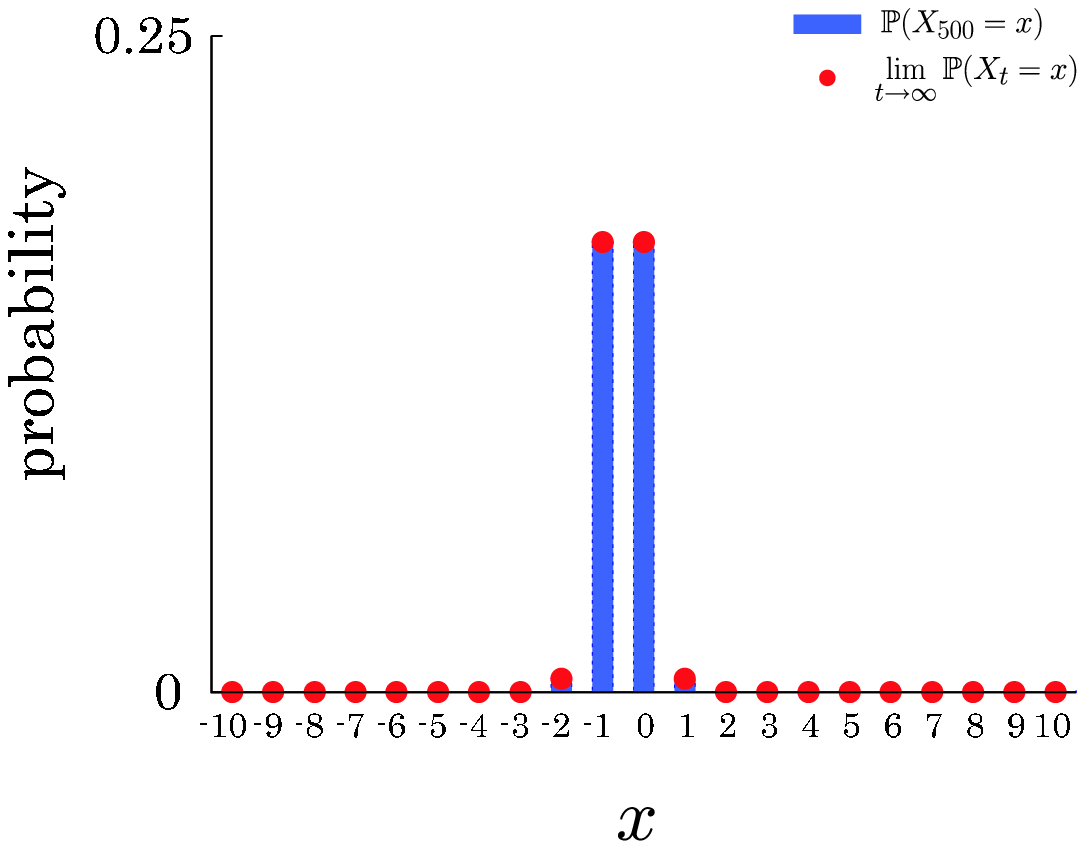}\\[2mm]
  (b) $\theta_1=\pi/4, \theta_2=\pi/4$
  \end{center}
 \end{minipage}
 \vspace{5mm}

 \begin{minipage}{50mm}
  \begin{center}
  \includegraphics[scale=0.4]{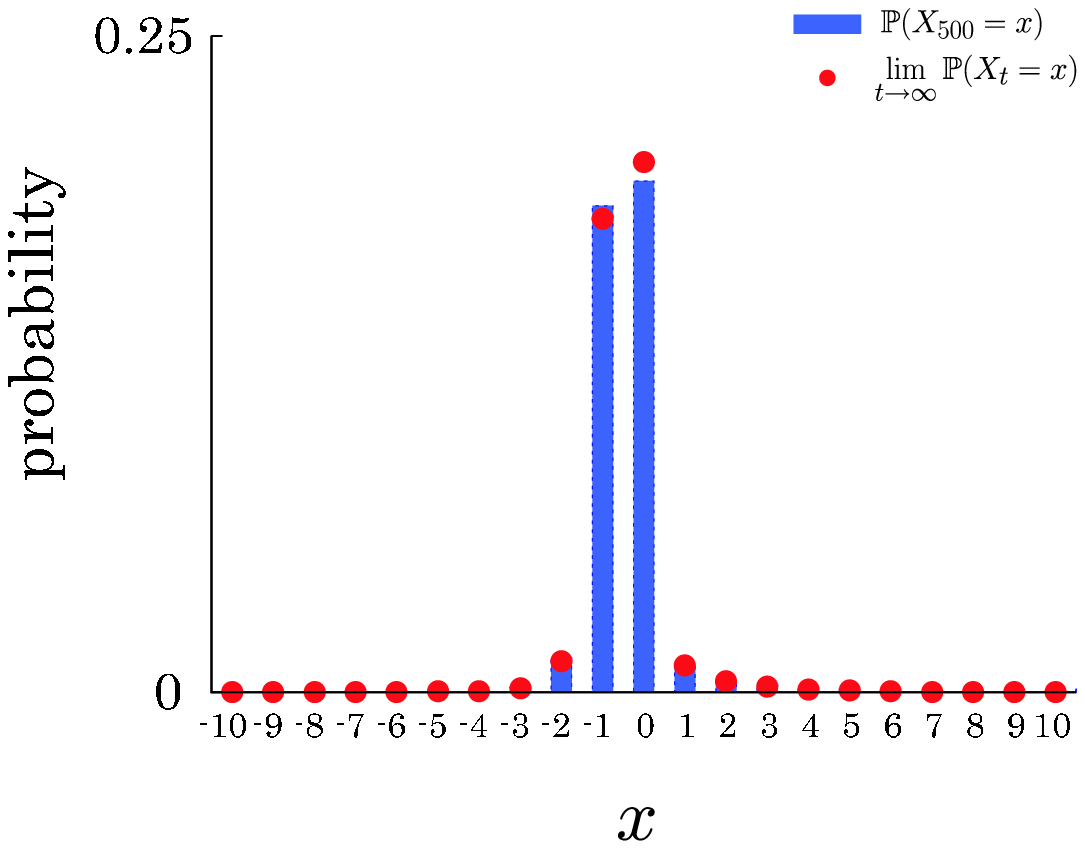}\\[2mm]
  (c) $\theta_1=\pi/6, \theta_2=\pi/4$
  \end{center}
 \end{minipage}
 \begin{minipage}{50mm}
  \begin{center}
 \includegraphics[scale=0.4]{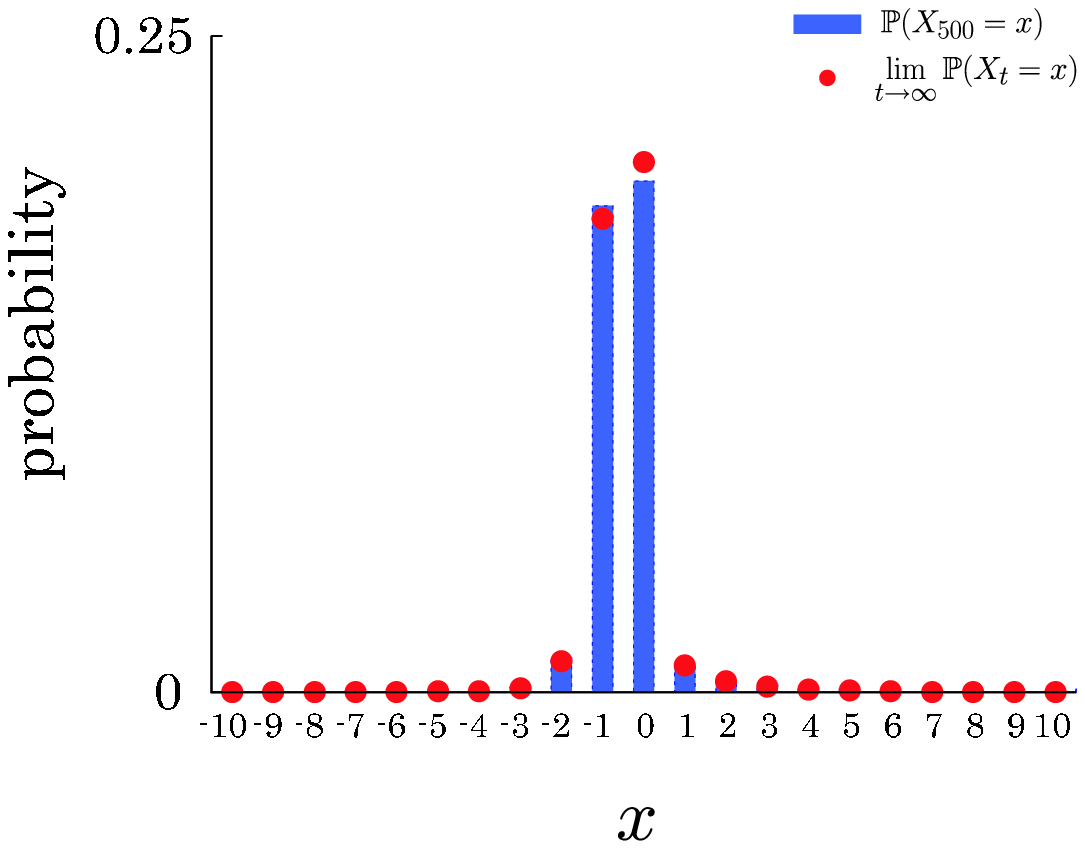}\\[2mm]
  (d) $\theta_1=\pi/4, \theta_2=\pi/6$
  \end{center}
 \end{minipage}
 \vspace{5mm}
 \caption{Given the initial state fixed at $q_0=q_1=q_2=q_3=1/2$, the long-time limit $\lim_{t\to\infty}\mathbb{P}(X_t=x)$ (red points) approximates to the probability at time $500$, $\mathbb{P}(X_{500}=x)$ (blue bars).}
 \label{fig:fig2}
\end{center}
\end{figure}

\begin{figure}[h]
\begin{center}
 \begin{minipage}{50mm}
  \begin{center}
  \includegraphics[scale=0.4]{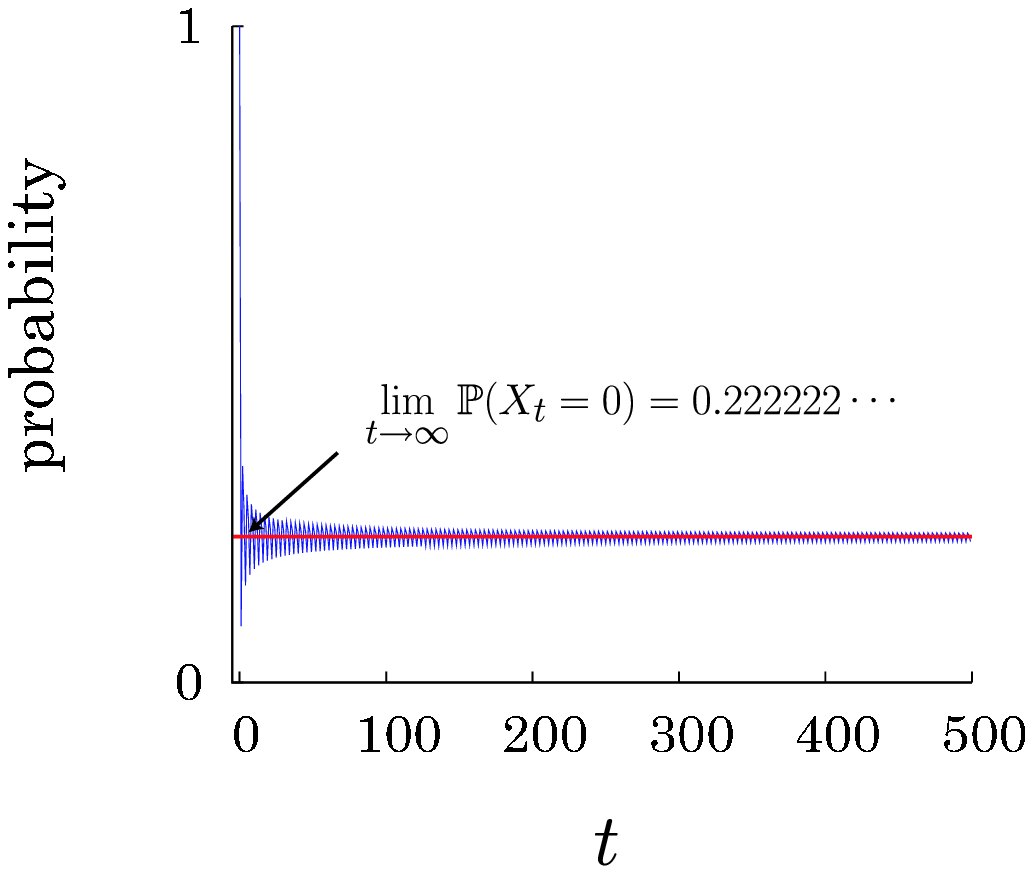}\\[2mm]
  (a) $\theta_1=\pi/6, \theta_2=\pi/6$
  \end{center}
 \end{minipage}
 \begin{minipage}{50mm}
  \begin{center}
 \includegraphics[scale=0.4]{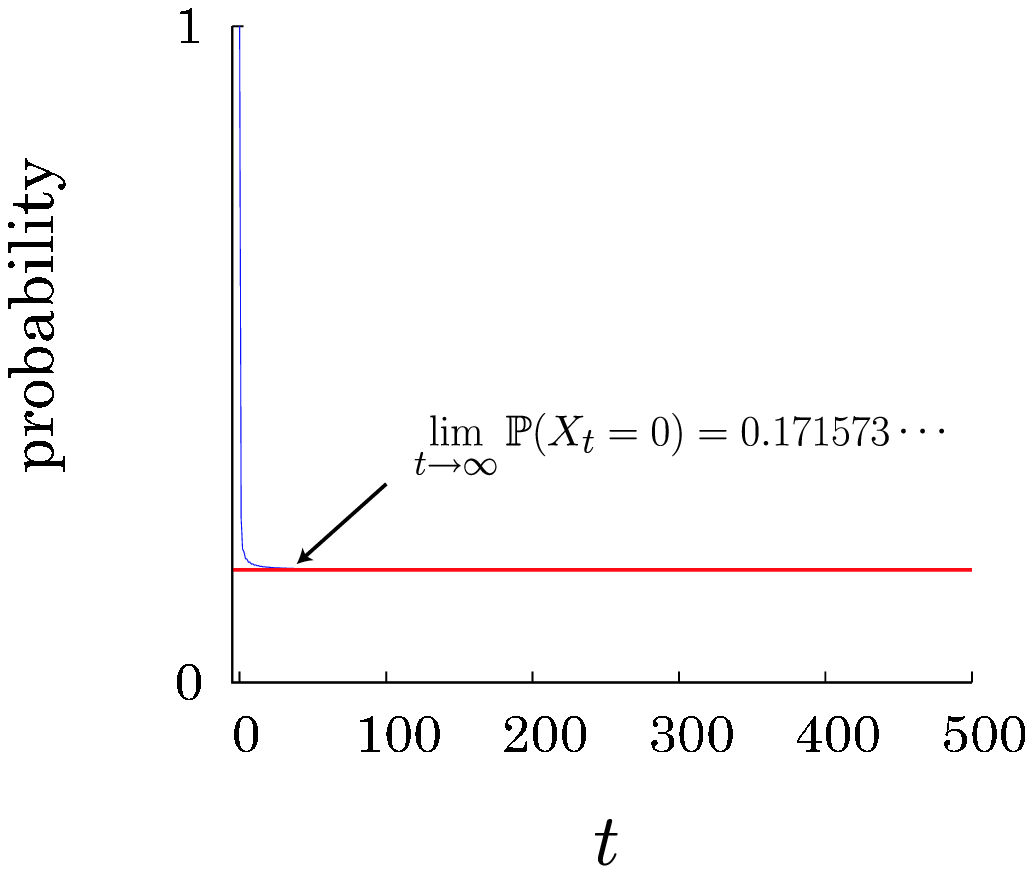}\\[2mm]
  (b) $\theta_1=\pi/4, \theta_2=\pi/4$
  \end{center}
 \end{minipage}
 \vspace{5mm}

 \begin{minipage}{50mm}
  \begin{center}
  \includegraphics[scale=0.4]{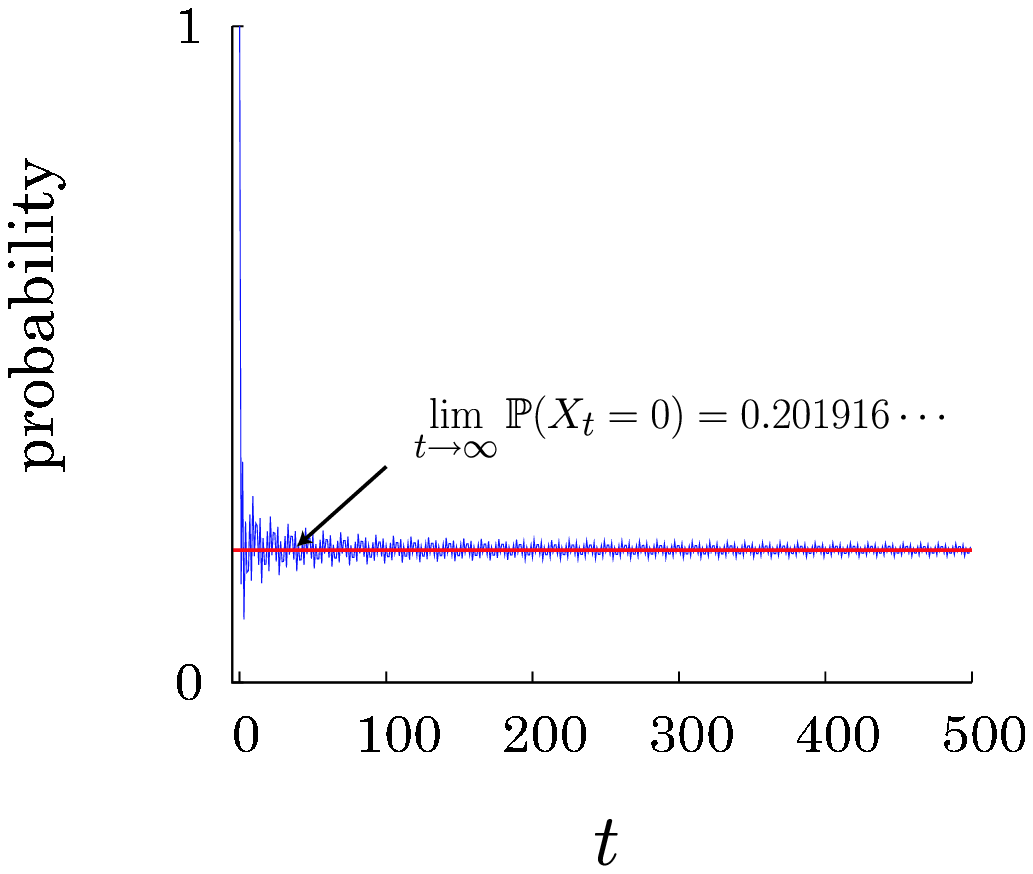}\\[2mm]
  (c) $\theta_1=\pi/6, \theta_2=\pi/4$
  \end{center}
 \end{minipage}
 \begin{minipage}{50mm}
  \begin{center}
 \includegraphics[scale=0.4]{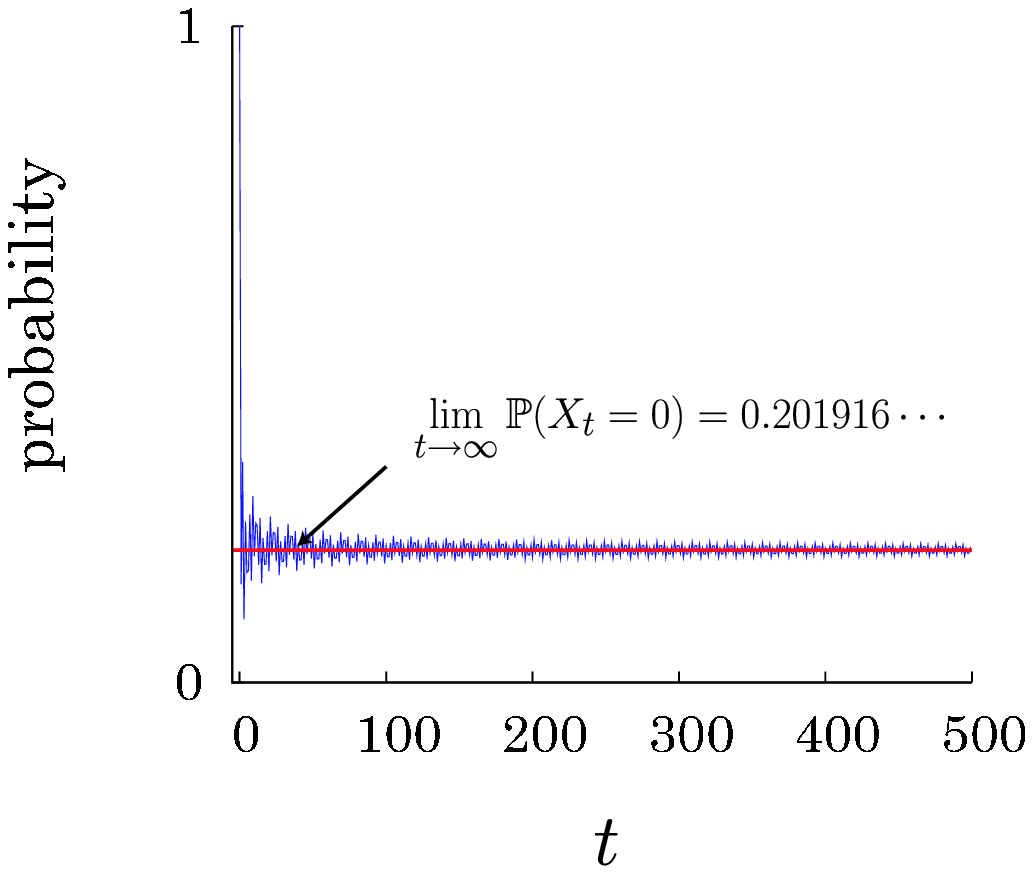}\\[2mm]
  (d) $\theta_1=\pi/4, \theta_2=\pi/6$
  \end{center}
 \end{minipage}
 \vspace{5mm}
 \caption{The probability that the quantum walker is observed at position $x=0$, $\mathbb{P}(X_t=0)$ (blue line), converges to the corresponding limit value, $\lim_{t\to\infty}\mathbb{P}(X_t=0)$ (red line), with $q_0=q_1=q_2=q_3=1/2$.}
 \label{fig:fig3}
\end{center}
\end{figure}

Before moving on to the proof of the limit theorem, we give in some figures below a comparison between some numerical simulations and the values asserted in the theorem for the probability $\mathbb{P}(X_t=x)$ at a large time $t$.
We see an almost perfect match, for a large value of $t$  and all values of $x$ in Fig.~\ref{fig:fig2}.
For varying values of $t$ , the probability at position $x=0$ is seen to converge nicely to the corresponding limit in Fig.~\ref{fig:fig3}.

The limit values of the probability distribution can be obtained by using Fourier analysis, a method introduced in the study of weak limit laws for quantum walks in 2004~\cite{GrimmettJansonScudo2004}.
For earlier use of the Fourier method in the study of quantum walks, see~\cite{NV,ABNVW}. For other ways to study the asymptotic behavior of quantum walks, see for instance~\cite{CIR}.

Now we give the proof of Theorem 1.
The Fourier transform of the quantum walk at time $t$,
\begin{equation}
 \ket{\hat \psi_t(k)}=\sum_{x\in\mathbb{Z}}e^{-ikx} \bra{x}\otimes\left(\sum_{j_1=0}^1\sum_{j_2=0}^1\ket{j_1j_2}\bra{j_1j_2}\right)\,\ket{\Psi_t},
\end{equation}
gets updated at time $t+1$, to the state
\begin{align}
 \ket{\hat \psi_{t+1}(k)}=&\left\{\Bigl(\hat U_1(k)\hat U_2(k)\Bigr)\otimes\Bigl(\hat U_2(k)\hat U_1(k)\Bigr)\right\}\ket{\hat \psi_t(k)}\nonumber\\
=&\left\{\Bigl(\hat U_1(k)\hat U_2(k)\Bigr)\otimes\Bigl(\hat U_2(k)\hat U_1(k)\Bigr)\right\}^t\ket{\phi},\label{eq:Fourier_evolution}
\end{align}
where
\begin{equation}
 \hat U_j(k)=\Bigl(e^{ik/2}\ket{0}\bra{0}+e^{-ik/2}\ket{1}\bra{1}\Bigr)\,U_j\quad (j=1,2).
\end{equation}
and $\ket{\phi}$ stands for the initial state $\ket{\hat \psi_{0}(k)}$.
Above we have made use of the identity
\begin{equation}
 \begin{bmatrix}
  e^{ik} & 0 & 0 & 0\\
  0 & 1 & 0 & 0\\
  0 & 0 & 1 & 0\\
  0 & 0 & 0 & e^{-ik}
 \end{bmatrix}
 =\begin{bmatrix}
   e^{ik/2} &0\\
   0 & e^{-ik/2}
  \end{bmatrix}
  \otimes
  \begin{bmatrix}
   e^{ik/2} &0\\
   0 & e^{-ik/2}
  \end{bmatrix}
\end{equation}
with
\begin{equation}
 \ket{0}=\begin{bmatrix}
	  1\\ 0
	 \end{bmatrix},\quad
	 \ket{1}=\begin{bmatrix}
		  0\\ 1
		 \end{bmatrix}.
\end{equation}
Expression~\eqref{eq:Fourier_evolution} has been derived from Eq.~\eqref{eq:time_evolution}.
Both unitary operations $\hat U_1(k)\hat U_2(k)$ and $\hat U_2(k)\hat U_1(k)$ have the same two eigenvalues
\begin{equation}
 \lambda_j(k)=c_1c_2\cos k+s_1s_2-i\,(-1)^j\sqrt{1-(c_1c_2\cos k+s_1s_2)^2}\quad (j=1,2).
\end{equation}
These eigenvalues for this pair of $2\times 2$ matrices, namely $\hat U_1(k)\hat U_2(k)$ and $\hat U_2(k)\hat U_1(k)$, give the eigenvalues of our 4-state QW $(\hat U_1(k)\hat U_2(k))\otimes (\hat U_2(k)\hat U_1(k))$ as $\lambda_1(k)\lambda_1(k),\lambda_1(k)\lambda_2(k),\lambda_2(k)\lambda_1(k),\lambda_2(k)\lambda_2(k)$.
From the relation $\lambda_2(k)=\overline{\lambda_1(k)}$, we have four eigenvalues for our QW, namely
\begin{equation}
 \lambda_1(k)^2,\, 1,\, 1,\, \lambda_2(k)^2.\label{eq:four_eigenvalues}
\end{equation}
Let $\ket{u_j(k)}\,(j\in\left\{1,2\right\})$ be the normalized eigenvector of the unitary operation $\hat U_1(k)\hat U_2(k)$ associated to the eigenvalue $\lambda_j(k)$.
Then we have a representation of the eigenvector,
\begin{align}
 \ket{u_j(k)}=&\frac{1}{\sqrt{N_j(k)}}\biggl[(c_1s_2 e^{ik}-s_1c_2)\ket{0}\nonumber\\
 &+i\left\{-c_1c_2\sin k-(-1)^j\sqrt{1-(c_1c_2\cos k+s_1s_2)^2}\right\}\ket{1}\biggr],
\end{align}
with the normalized factor
\begin{align}
 N_j(k)=&2\sqrt{1-(c_1c_2\cos k+s_1s_2)^2}\nonumber\\
 &\times\left\{\sqrt{1-(c_1c_2\cos k+s_1s_2)^2}+(-1)^j c_1c_2\sin k\right\}.
\end{align}
On the other hand, the normalized eigenvector, represented by $\ket{v_j(k)}\,(j\in\left\{1,2\right\})$, of the unitary operation $\hat U_2(k)\hat U_1(k)$ associated to the eigenvalue $\lambda_j(k)$ has a representation 
\begin{align}
 \ket{v_j(k)}=&\frac{1}{\sqrt{N_j(k)}}\biggl[(s_1c_2 e^{ik}-c_1s_2)\ket{0}\nonumber\\
 &+i\left\{-c_1c_2\sin k-(-1)^j\sqrt{1-(c_1c_2\cos k+s_1s_2)^2}\right\}\ket{1}\biggr],
\end{align}
We should note that $N_1(k)N_2(k)$ can be arranged in the form
\begin{align}
 N_1(k)N_2(k)=&\frac{c_1^3c_2^3s_1s_2}{(e^{ik})^3}
 \left\{e^{ik}-\frac{(1+s_1)(1-s_2)}{c_1c_2}\right\}
 \left\{e^{ik}-\frac{(1-s_1)(1+s_2)}{c_1c_2}\right\}\nonumber\\
 &\times\left\{e^{ik}+\frac{(1+s_1)(1+s_2)}{c_1c_2}\right\}
 \left\{e^{ik}+\frac{(1-s_1)(1-s_2)}{c_1c_2}\right\}\nonumber\\
 &\times\left(e^{ik}-\frac{s_1c_2}{c_1s_2}\right)
 \left(e^{ik}-\frac{c_1s_2}{s_1c_2}\right)\nonumber\\
=&\frac{c_1^3c_2^3s_1s_2}{(e^{ik})^3}
 (e^{ik}-\nu_1)(e^{ik}-\nu_1^{-1})(e^{ik}-\nu_2)(e^{ik}-\nu_2^{-1})\nonumber\\
 &\times\left(e^{ik}-\frac{s_1c_2}{c_1s_2}\right)
 \left(e^{ik}-\frac{c_1s_2}{s_1c_2}\right).
\end{align}
Therefore the unitary operation $(\hat U_1(k)\hat U_2(k))\otimes (\hat U_2(k)\hat U_1(k))$, has four normalized eigenvectors, represented by $\ket{w_{j_1j_2}(k)}\,(j_1,j_2=1,2)$, associated to the eigenvalues $\lambda_{j_1}(k)\lambda_{j_2}(k)$,
\begin{equation}
 \ket{w_{j_1j_2}(k)}=\ket{u_{j_1}(k)}\otimes\ket{v_{j_2}(k)}.
\end{equation}
Decomposing the initial state in the eigenspace spanned by the orthonormal basis $\left\{\ket{w_{11}}, \ket{w_{12}}, \ket{w_{21}}, \ket{w_{22}}\right\}$,
\begin{equation}
 \ket{\phi}=\sum_{j_1=1}^2\sum_{j_2=1}^2 \braket{w_{j_1j_2}(k)|\phi}\,\ket{w_{j_1j_2}(k)},
\end{equation}
we can also express the Fourier transform as follows
\begin{align}
 \ket{\hat\psi_t(k)}=&\left\{\Bigl(\hat U_1(k)\hat U_2(k)\Bigr)\otimes\Bigl(\hat U_2(k)\hat U_1(k)\Bigr)\right\}^t\ket{\phi}\nonumber\\
 =&\sum_{j_1=1}^2\sum_{j_2=1}^2 \lambda_{j_1}(k)^t\lambda_{j_2}(k)^t\braket{w_{j_1j_2}(k)|\phi}\,\ket{w_{j_1j_2}(k)}\nonumber\\
 =&\braket{w_{12}(k)|\phi}\,\ket{w_{12}(k)}+\braket{w_{21}(k)|\phi}\,\ket{w_{21}(k)}\nonumber\\
 &+\lambda_1(k)^{2t}\braket{w_{11}(k)|\phi}\,\ket{w_{11}(k)}+\lambda_2(k)^{2t}\braket{w_{22}(k)|\phi}\,\ket{w_{22}(k)}.
\end{align}
Using the expression above, an application of the Riemann--Lebesgue lemma, see~\cite{DMc}, page 101, gives in the limit of large $t$ a useful expression, namely  
\begin{align}
 &\lim_{t\to\infty}\biggl(\bra{x}\otimes\sum_{j_1=0}^1\sum_{j_2=0}^1\ket{j_1j_2}\bra{j_1j_2}\biggr)\ket{\Psi_t}\nonumber\\
 =&\lim_{t\to\infty}\int_{-\pi}^\pi e^{ikx}\ket{\hat\psi_t(k)}\,\frac{dk}{2\pi}\nonumber\\
 =&\int_{-\pi}^\pi e^{ikx} \Bigl(\braket{w_{12}(k)|\phi}\,\ket{w_{12}(k)}+\braket{w_{21}(k)|\phi}\,\ket{w_{21}(k)}\Bigr)\,\frac{dk}{2\pi}.
\end{align}
The integral can be computed to be of the form 
\begin{align}
 &\int_{-\pi}^\pi e^{ikx} \Bigl(\braket{w_{12}(k)|\phi}\,\ket{w_{12}(k)}+\braket{w_{21}(k)|\phi}\,\ket{w_{21}(k)}\Bigr)\,\frac{dk}{2\pi}\nonumber\\
 =&\Bigl(\eta_1(x)q_0-\eta_3(x)q_1-\eta_3(-x-1)q_2-\eta_1(x+1)q_3\Bigr)\ket{00}\nonumber\\
 &+\Bigl(-\eta_3(-x)q_0+(\delta_{0x}-\eta_1(x))q_1+\eta_2(-x)q_2+\eta_3(-x-1)q_3\Bigr)\ket{01}\nonumber\\
 &+\Bigl(-\eta_3(x-1)q_0+\eta_2(x)q_1+(\delta_{0x}-\eta_1(x))q_2+\eta_3(x)q_3\Bigr)\ket{10}\nonumber\\
 &+\Bigl(-\eta_1(x-1)q_0+\eta_3(x-1)q_1+\eta_3(-x)q_2+\eta_1(x)q_3\Bigr)\ket{11},
\end{align}
from which the statement in expression \eqref{eq:limit_measure} follows.

\section{Convergence in distribution}
\label{sec:convergence_in_distribution}
Besides enabling us to study the limit of the probability $\mathbb{P}(X_t=x)\,(x\in\mathbb{Z})$, the use of Fourier analysis enables us to get our hands on the distribution $\mathbb{P}(X_t/t\leq x)\,(x\in\mathbb{R})$ for large values of $t$.

\begin{thm}
\label{th:limit_theorem_2}
Let $m$ and $M$ $\in\{1,2\}$ be the subscripts such that $|c_m|=\min\left\{|c_1|,\,|c_2|\right\}$ and $|c_M|=\max\left\{|c_1|,\,|c_2|\right\}$.
For a real number $x\,(\neq 0)$, we have convergence in distribution, namely
\begin{align}
 &\lim_{t\to\infty}\mathbb{P}\left(\frac{X_t}{t}\leq x\right)\nonumber\\
=&\int_{-\infty}^x \biggl\{\Delta\,\delta_0(y)\,+\,\frac{|s_m|\,(d_0+d_1y+d_2y^2)}{\,\pi(4-y^2)\sqrt{4c_m^2-y^2}\,}\,I_{\left(-2|c_m|,\,2|c_m|\right)}(y)\biggr\}\,dy,\label{eq:limit_distribution}
\end{align}
where
\begin{align}
 \Delta=&|q_1|^2+|q_2|^2+\Bigl(|q_0|^2-|q_1|^2-|q_2|^2+|q_3|^2\Bigr)\eta_1(0)\label{eq:delta_Th2}\nonumber\\
 &-2\Bigl\{\Re(q_0\overline{q_3})\eta_1(1)-\Re(q_1\overline{q_2})\eta_2(0)\nonumber\\
 &+\Re(q_0\overline{q_1}-q_2\overline{q_3})\eta_3(0)-\Re(q_0\overline{q_2}-q_1\overline{q_3})\eta_3(-1)\Bigr\},
\end{align}
\begin{align}
 d_0=&2\left\{1+\frac{2c_m^2s_M}{\,c_1c_2s_m\,}\,\Re(q_1\overline{q_2}-q_0\overline{q_3})\right\},\\
 d_1=&-2\left\{|q_0|^2-|q_3|^2+\frac{s_1}{\,c_1\,}\,\Re(q_0\overline{q_1}+q_2\overline{q_3})+\frac{s_2}{\,c_2\,}\,\Re(q_0\overline{q_2}+q_1\overline{q_3})\right\},\\
 d_2=&\frac{1}{\,2\,}\Bigl(|q_0|^2-|q_1|^2-|q_2|^2+|q_3|^2\Bigr)\nonumber\\
 &+\frac{s_1}{\,c_1\,}\,\Re(q_0\overline{q_1}-q_2\overline{q_3})+\frac{s_2}{\,c_2\,}\,\Re(q_0\overline{q_2}-q_1\overline{q_3})\nonumber\\
 &+\frac{s_1s_2}{\,c_1c_2\,}\,\Re(q_0\overline{q_3}+q_1\overline{q_2})+\frac{s_M}{\,c_1c_2s_m\,}\,\Re(q_0\overline{q_3}-q_1\overline{q_2}),
\end{align}
\begin{align}
 I_{\left(\,-2|c_m|,\,2|c_m|\right)}(x)=&\left\{\begin{array}{ll}
     1&(\,-2|c_m| < x < 2|c_m|\,)\\[1mm]
					       0& (\mbox{otherwise})
					      \end{array}\right..
\end{align}
The function $\delta_0(x)$ denotes a Dirac $\delta$-function at the origin $x=0$.
The notation $\Re(z)$ denotes the real part of a complex number $z$.
Finally, the notations $s_m$ and $s_M$ are defined as $s_m=\sin\theta_m$ and $s_M=\sin\theta_M$, with $m$ and $M$ spelled out above.
\end{thm}

This kind of limit distribution was also obtained for a quantum walk with a doubly entangled coin operation~\cite{Li}, and our limit distribution agrees with the previous result when $\theta_1=\theta_2$.

\begin{figure}[h]
\begin{center}
 \begin{minipage}{50mm}
  \begin{center}
  \includegraphics[scale=0.4]{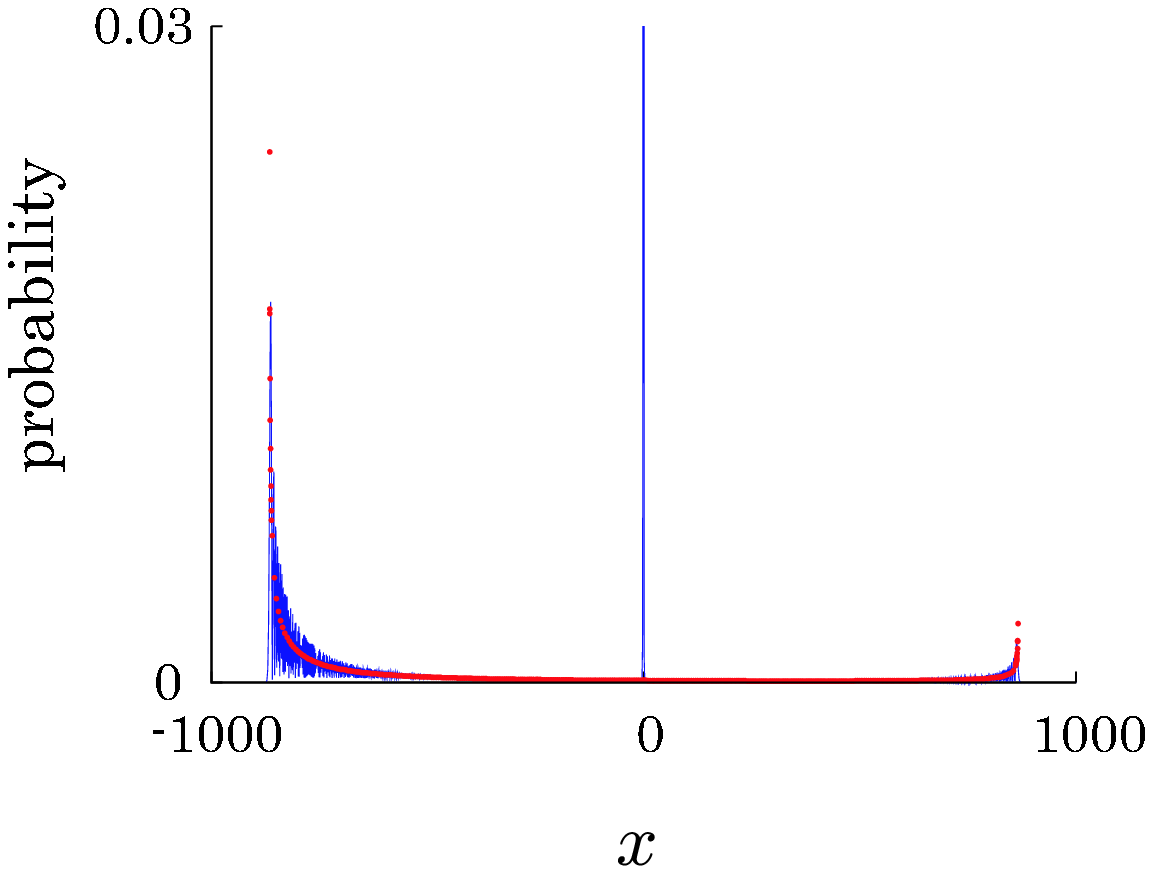}\\[2mm]
  (a) $\theta_1=\pi/6, \theta_2=\pi/6$
  \end{center}
 \end{minipage}
 \begin{minipage}{50mm}
  \begin{center}
 \includegraphics[scale=0.4]{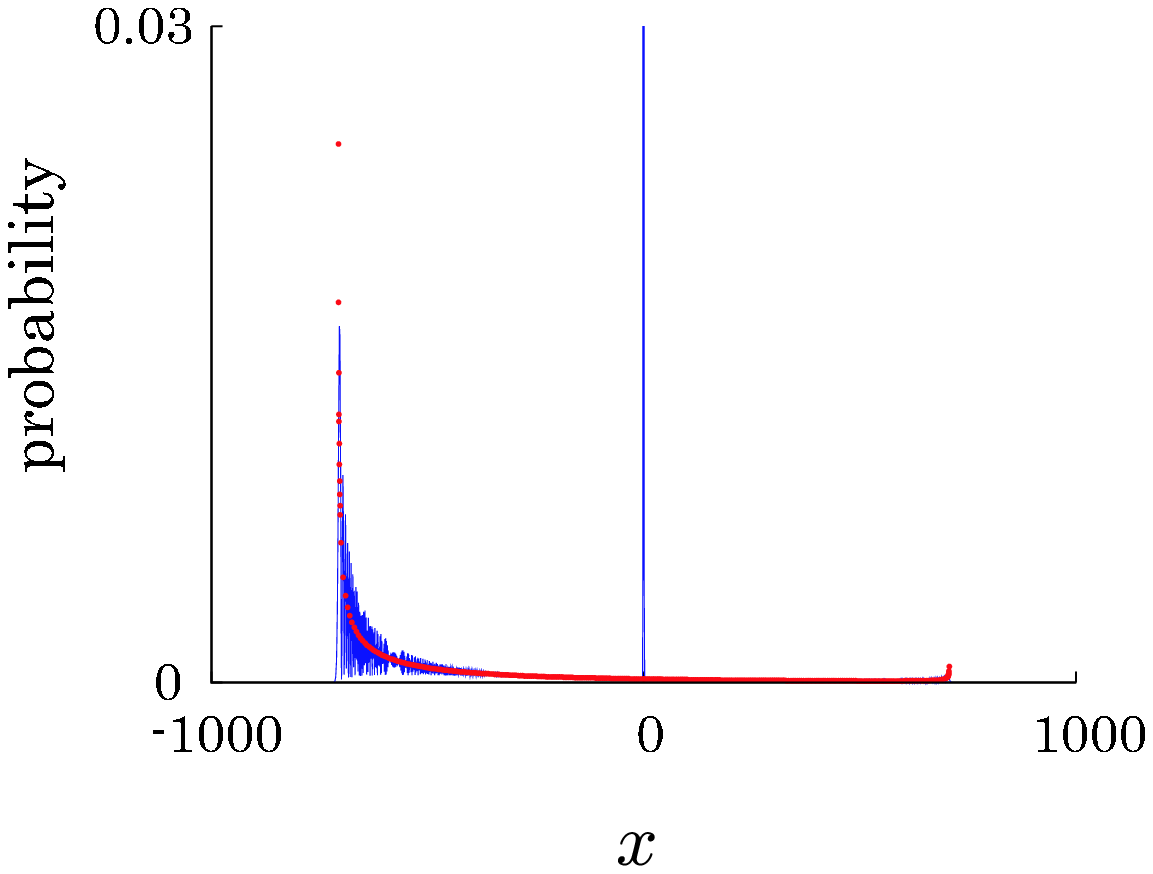}\\[2mm]
  (b) $\theta_1=\pi/4, \theta_2=\pi/4$
  \end{center}
 \end{minipage}
 \vspace{5mm}

 \begin{minipage}{50mm}
  \begin{center}
  \includegraphics[scale=0.4]{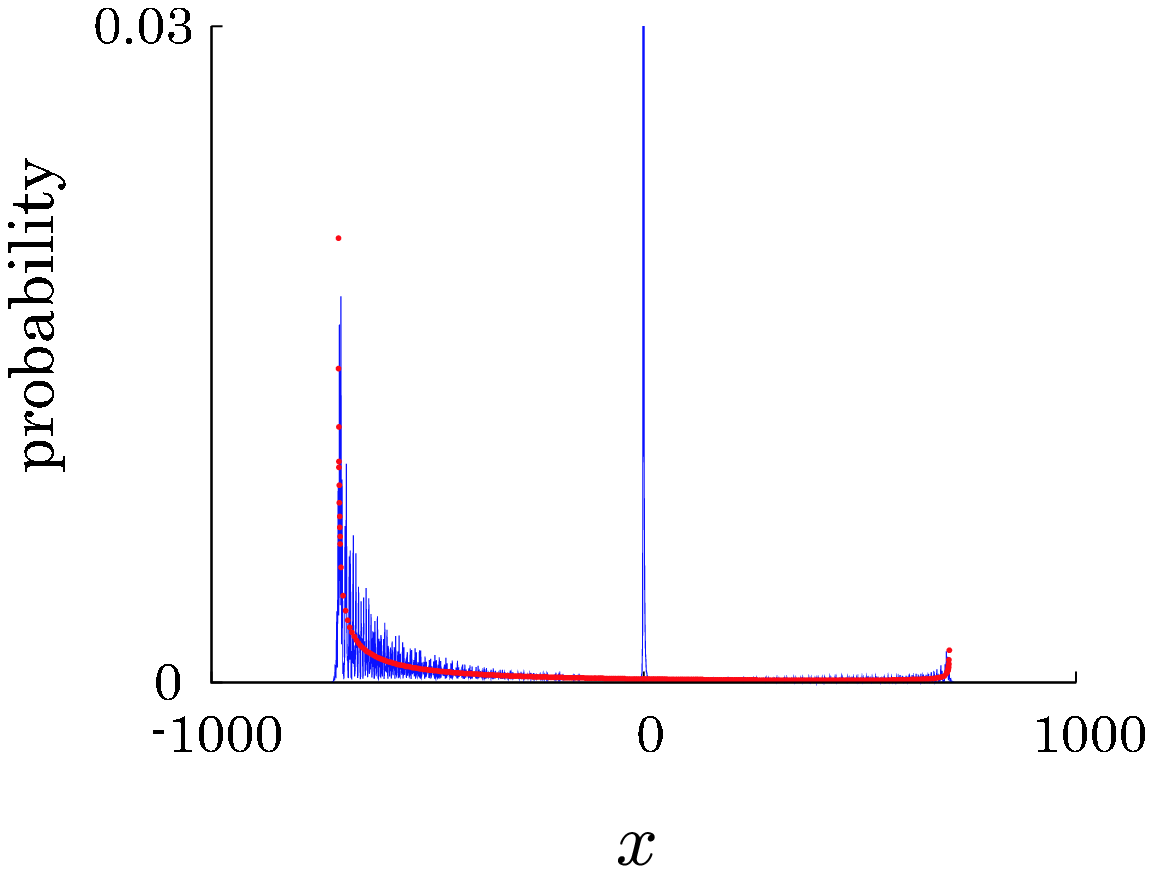}\\[2mm]
  (c) $\theta_1=\pi/6, \theta_2=\pi/4$
  \end{center}
 \end{minipage}
 \begin{minipage}{50mm}
  \begin{center}
 \includegraphics[scale=0.4]{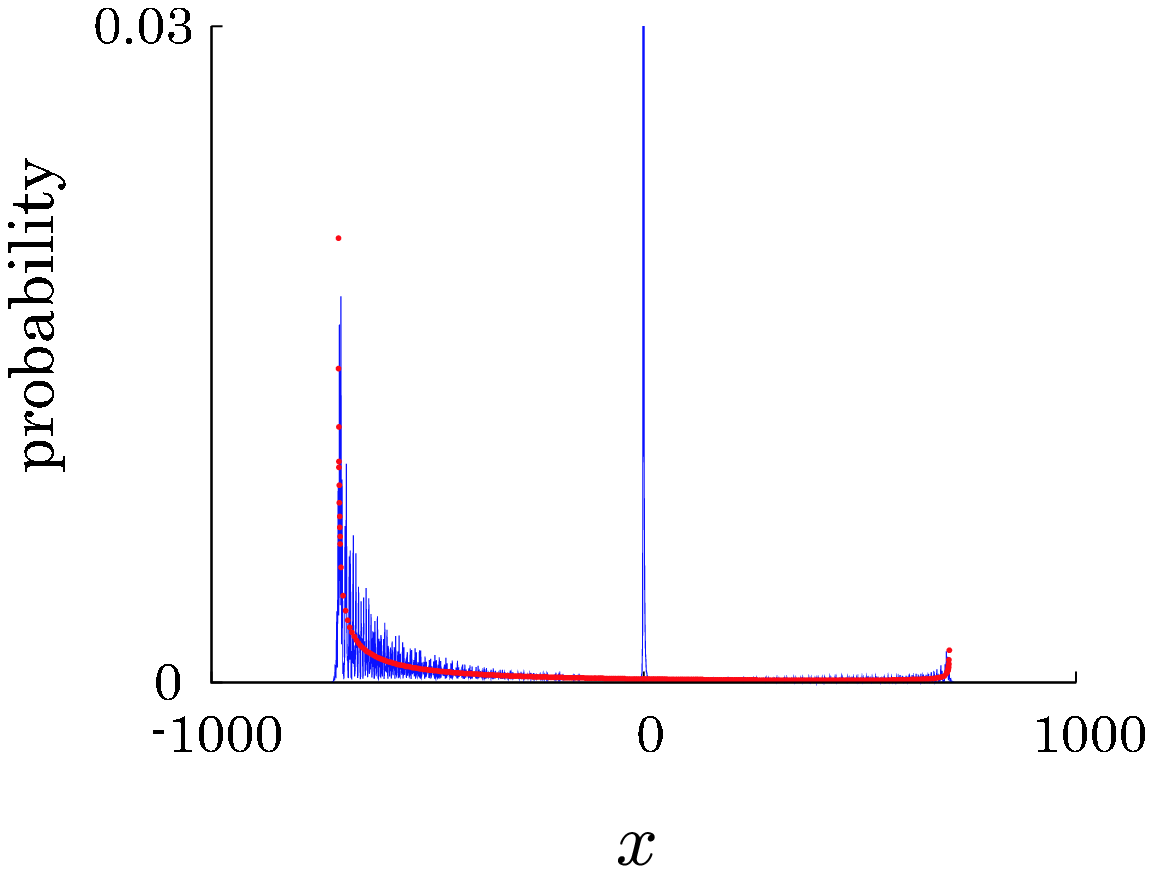}\\[2mm]
  (d) $\theta_1=\pi/4, \theta_2=\pi/6$
  \end{center}
 \end{minipage}
 \vspace{5mm}
 \caption{An approximation obtained from the continuous-part of the limit density function (red points) reproduces the ballistic behavior (delocalizing part) of the quantum walker in distribution $\mathbb{P}(X_{500}=x)$ (blue line) with $q_0=q_1=q_2=q_3=1/2$}
 \label{fig:fig4}
\end{center}
\end{figure}

The continuous part of the limit density function
\begin{equation}
 f(x)=\frac{|s_m|\,(d_0+d_1x+d_2x^2)}{\,\pi(4-x^2)\sqrt{4c_m^2-x^2}\,}\,I_{\left(-2|c_m|,\,2|c_m|\right)}(x)
\end{equation}
can yield an approximation to the probability $\mathbb{P}(X_t=x)$ except for localization.
We give a comparison between $\mathbb{P}(X_t=x)$ and the approximation $1/t\cdot f(x/t)$ in Fig.~\ref{fig:fig4}, and find that the approximation reproduces the delocalizing part of probability distribution.

Now we move into the proof of Theorem 2.
Our aim is to study the limit of the $r$-th moment, $\lim_{t\to\infty}\mathbb{E}[(X_t/t)^r]\,(r=0,1,2,\ldots)$, which will yield convergence in distribution to an appropriate law.
With the notation $D=i\,(d/dk)$, the representation of the $r$-th moment in the form
\begin{equation}
 \mathbb{E}(X_t^r)=\int_{-\pi}^\pi \bra{\hat\psi_t(k)}\Bigl(D^r\ket{\hat\psi_t(k)}\Bigr)\,\frac{dk}{2\pi},
\end{equation}
connects the $r$-th moment to the Fourier transform $\ket{\hat\psi_t(k)}$.
Arranging the $r$-th derivative in the following fashion,
\begin{align}
 &\frac{d^r}{dk^r}\ket{\hat\psi_t(k)}\nonumber\\
 =&(t)_r\Biggl[\sum_{j_1=1}^2\sum_{j_2=1}^2 \Bigl(\lambda_{j_1}(k)\lambda_{j_2}(k)\Bigr)^{t-r}\left\{\Bigl(\lambda_{j_1}(k)\lambda_{j_2}(k)\Bigr)'\right\}^r\nonumber\\
 &\times\braket{w_{j_1j_2}(k)|\phi}\ket{w_{j_1j_2}(k)}\Biggr]+\sum_{j_1=0}^1\sum_{j_2=0}^1 O(t^{r-1})\ket{j_1j_2},
\end{align}
we have
\begin{align}
 &\bra{\hat\psi_t(k)}\Bigl(D^r\ket{\hat\psi_t(k)}\Bigr)\nonumber\\
 =&(t)_r\left[\sum_{j_1=1}^2\sum_{j_2=1}^2 \left\{\frac{i\Bigl(\lambda_{j_1}(k)\lambda_{j_2}(k)\Bigr)'}{\lambda_{j_1}(k)\lambda_{j_2}(k)}\right\}^r\Bigl|\braket{w_{j_1j_2}(k)|\phi}\Bigr|^2\right]+O(t^{r-1}),
\end{align}
where $(t)_r=\Pi_{j=1}^t(t-j+1)$.
The $r$-th moment $\mathbb{E}[(X_t/t)^r]$ converges to an integral form as $t\to\infty$,
\begin{align}
 &\lim_{t\to\infty}\mathbb{E}\left[\left(\frac{X_t}{t}\right)^r\right]
=\lim_{t\to\infty}\frac{\mathbb{E}(X_t^r)}{t^r}\nonumber\\
=&\lim_{t\to\infty}\frac{1}{t^r}\int_{-\pi}^\pi \bra{\hat\psi_t(k)}\Bigl(D^r\ket{\hat\psi_t(k)}\Bigr)\,\frac{dk}{2\pi}\nonumber\\
=&\int_{-\pi}^\pi \sum_{j_1=1}^2\sum_{j_2=1}^2 \left\{\frac{i\Bigl(\lambda_{j_1}(k)\lambda_{j_2}(k)\Bigr)'}{\lambda_{j_1}(k)\lambda_{j_2}(k)}\right\}^r\Bigl|\braket{w_{j_1j_2}(k)|\phi}\Bigr|^2\,\frac{dk}{2\pi}.\nonumber\\
\end{align}
Of the four terms in this last sum, two of them come from the eigenvalue $\equiv 1$ of the operator $\Bigl(\hat U_1(k)\hat U_2(k)\Bigr)\otimes\Bigl(\hat U_2(k)\hat U_1(k)\Bigr)$, while the other two come from the $k$ dependent eigenvalues $\lambda_1(k)^2, \lambda_2(k)^2$, see \eqref{eq:four_eigenvalues}.
The first two terms give the value one for the expression $\left\{i(\lambda_{j_1}(k)\lambda_{j_2}(k))'/\lambda_{j_1}(k)\lambda_{j_2}(k)\right\}^r$ when $r=0$ and zero otherwise, and then using the convention
\begin{equation}
 0^r=\left\{\begin{array}{ll}
      1& (r=0)\\
	     0& (\mbox{otherwise})
	    \end{array}\right.,
\end{equation}
the sum of the four terms can be written as
\begin{align}
&\int_{-\pi}^\pi 0^r\left(\Bigl|\braket{w_{12}(k)|\phi}\Bigr|^2+\Bigl|\braket{w_{21}(k)|\phi}\Bigr|^2\right)\,\frac{dk}{2\pi}\nonumber\\
&+\int_{-\pi}^\pi \left(\frac{2i\lambda_1'(k)}{\lambda_1(k)}\right)^r\Bigl|\braket{w_{11}(k)|\phi}\Bigr|^2\,\frac{dk}{2\pi}+\int_{-\pi}^\pi \left(\frac{2i\lambda_2'(k)}{\lambda_2(k)}\right)^r\Bigl|\braket{w_{22}(k)|\phi}\Bigr|^2\,\frac{dk}{2\pi}\nonumber\\
=&0^r\cdot \Delta
+\int_{-\pi}^\pi \left(\frac{2i\lambda_1'(k)}{\lambda_1(k)}\right)^r\Bigl|\braket{w_{11}(k)|\phi}\Bigr|^2\,\frac{dk}{2\pi}\nonumber\\
&+\int_{-\pi}^\pi \left(\frac{2i\lambda_2'(k)}{\lambda_2(k)}\right)^r\Bigl|\braket{w_{22}(k)|\phi}\Bigr|^2\,\frac{dk}{2\pi}\nonumber\\
=&\int_{-\infty}^\infty x^r\cdot \Delta\,\delta_0(x)\,dx\nonumber\\
&+\int_{-\pi}^\pi \left(\frac{2i\lambda_1'(k)}{\lambda_1(k)}\right)^r\Bigl|\braket{w_{11}(k)|\phi}\Bigr|^2\,\frac{dk}{2\pi}+\int_{-\pi}^\pi \left(\frac{2i\lambda_2'(k)}{\lambda_2(k)}\right)^r\Bigl|\braket{w_{22}(k)|\phi}\Bigr|^2\,\frac{dk}{2\pi},
\end{align}
where
\begin{align}
 \Delta=&\int_{-\pi}^\pi \left(\Bigl|\braket{w_{12}(k)|\phi}\Bigr|^2+\Bigl|\braket{w_{21}(k)|\phi}\Bigr|^2\right)\,\frac{dk}{2\pi}\nonumber\\
 =&|q_1|^2+|q_2|^2+\Bigl(|q_0|^2-|q_1|^2-|q_2|^2+|q_3|^2\Bigr)\eta_1(0)\nonumber\\
 &-2\Bigl\{\Re(q_0\overline{q_3})\eta_1(1)-\Re(q_1\overline{q_2})\eta_2(0)\nonumber\\
 &+\Re(q_0\overline{q_1}-q_2\overline{q_3})\eta_3(0)-\Re(q_0\overline{q_2}-q_1\overline{q_3})\eta_3(-1)\Bigr\}.
\end{align}
Since we have
\begin{equation}
 \frac{2i\lambda'_j(k)}{\lambda_j(k)}=(-1)^j\frac{2c_1c_2\sin k}{\sqrt{1-(c_1c_2\cos k+s_1s_2)^2}}\quad (j=1,2),
\end{equation}
the substitution $2i\lambda'_j(k)/\lambda_j(k)=x\,(j=1,2)$ makes another integral representation possible, namely
\begin{align}
 &\int_{-\pi}^\pi \left(\frac{2i\lambda_1'(k)}{\lambda_1(k)}\right)^r\Bigl|\braket{w_{11}(k)|\phi}\Bigr|^2\,\frac{dk}{2\pi}+\int_{-\pi}^\pi \left(\frac{2i\lambda_2'(k)}{\lambda_2(k)}\right)^r\Bigl|\braket{w_{22}(k)|\phi}\Bigr|^2\,\frac{dk}{2\pi}\nonumber\\
 =&\int_{-2|c_m|}^{2|c_m|} x^r \cdot\frac{|s_m|\,(d_0+d_1x+d_2x^2)}{\,\pi(4-x^2)\sqrt{4c_m^2-x^2}\,}\,dx\nonumber\\
=&\int_{-\infty}^\infty x^r \cdot\frac{|s_m|\,(d_0+d_1x+d_2x^2)}{\,\pi(4-x^2)\sqrt{4c_m^2-x^2}\,}\,I_{\left(-2|c_m|,\,2|c_m|\right)}(x)\,dx.
\end{align}
As a result, we have an integral form for the $r$-th moment, namely
\begin{align}
 &\lim_{t\to\infty}\mathbb{E}\left[\left(\frac{X_t}{t}\right)^r\right]\nonumber\\
=&\int_{-\infty}^\infty x^r\cdot \biggl\{\Delta\,\delta_0(x)\,+\,\frac{|s_m|\,(d_0+d_1x+d_2x^2)}{\,\pi(4-x^2)\sqrt{4c_m^2-x^2}\,}\,I_{\left(-2|c_m|,\,2|c_m|\right)}(x)\biggr\}\,dx,
\end{align}
and this guarantees the convergence in distribution asserted in Theorem~\ref{th:limit_theorem_2}.
\bigskip

It may be instructive to observe that in going from an integral in the variable $k$ to one in the variable $x$ we have used the equivalence, for nice functions, of two definitions of an integral due to Riemann and Lebesgue respectively.
More explicitly if we have a random variable $Y$ such that for every $r=0,1,2,\ldots$ we have
\begin{equation}
 \mathbb{E}(Y^r)=\int_{-\pi}^\pi h^r(k)\,p(k)\,dk,
\end{equation}
we get for real $\lambda$
\begin{equation}
 \mathbb{E}(e^{i\lambda Y})=\int_{-\pi}^\pi e^{i\lambda h(k)}\,p(k)\,dk.
\end{equation}
The left hand side is by definition
\begin{equation}
 \int e^{i\lambda x}\,d(\mbox{Prob}\,Y\leq x).
\end{equation}
On the other hand, the integral on the right hand side of the previous expression can be thought as the limit of Riemann sums
\begin{equation}
 \sum e^{i\lambda h(k'_j)}\,p(k'_j)\,(\Delta k)_j,
\end{equation}
when the partitions on the $k$ axis get finer and finer. This is Riemann's recipe.
But if, following Lebesgue's recipe, the partitions are done on ``other axis'', denoted here by $x$, the approximating sums will look like
\begin{equation}
 \sum e^{i\lambda x'_j}\,\times\, (\mbox{measure of those}\quad k\quad\mbox{with}\quad x_j\leq p(k)\leq x_{j+1}),
\end{equation}
and by refining these partitions we get in the limit the integral
\begin{equation}
 \int e^{i\lambda x}\,ds(x),
\end{equation}
where $s(x)\equiv \int_{p(k)\leq x}^x\,dk$.
Now a simple Fourier inversion gives the general form of the identity between two integrals that has been given above.
This kind of classical argument has been exposed in more detail in~\cite{DMc}, section 1.1, and is used in~\cite{Ko} to relate the results in~\cite{GrimmettJansonScudo2004} to earlier results of N.~Konno.
For a fuller discussion of weak convergence of measures, see Billingsley~\cite{Bi}.

\section{Application to an analytical study of a Parrondo type paradox}
\label{sec:application}

Nowadays one uses the term ``Parrondo paradox'' to refer to a situation where a combination of two winning games results in a losing one, or vice versa.
There are other such surprises in the world of stochastic and/or quantum processes, such as the ``Simpson paradox'', see~\cite{Sel} and its many useful references.
Parrondo's has a very distinguished pedigree originating in the famous {\bf Lectures in Physics} by Richard Feynman,~\cite{RF}.
J. Parrondo and R. Espagnol, wrote a thermodynamics paper,~\cite{PE}, questioning some of the arguments in one of Feynman's lectures, and this was quickly converted into a few nontechnical  papers, such as~\cite{PD}, accessible to a reader without the thermodynamical background of the first paper.
This very interesting topic, whose analysis requires only some basic linear algebra, has made its way into the textbook literature, see~\cite{SE}, page 155.
See also~\cite{HA1}.

Some people have tried to find quantum analogs of the classical situations alluded to above.
For a small selection see~\cite{meyer,MeB,flitney,HA,ChB,FA,LZG}.
These results tend to show that although a paradox may show up initially, if one waits long enough it disappears.
For variants of the original paradox see~\cite{pejicthesis,GP}.
This is such an alluring topic that one can expect several people to come up with different versions of this phenomenon.
The most recent efforts in this direction appear to be those reported in by J. Rajendran and C. Benjamin~\cite{RajendranBenjamin2018}.

\bigskip

One of the goals of our paper is to use the limit theorems given above to compute analytically some of the quantities obtained by these authors by means of very good numerical simulations.
We start by noticing a small point: our individual coins $U_1$ and $U_2$ are much simpler than the ones used in~\cite{RajendranBenjamin2018}.
Our limit theorems concern 2-period time-dependent walks, concretely the one obtained by iterating
\begin{equation}
 SY SX.
\end{equation}
One can obtain limit theorems for time-independent walks such as the one obtained by iterating $SX$.
This would involve two coins $U_1,U_2$ and the detailed results will be the purpose of a separate paper.
These points above are mentioned since in~\cite{RajendranBenjamin2018} the authors are comparing a 2-period time-dependent and a time-independent quantum walk, each based on two coins.
The first one agrees conceptually with ours, it is given by iterating $SY SX$ and it results in the sequence $U_1 U_2 U_1 U_2\cdots$ for one coin and the sequence $U_2 U_1 U_2 U_1\cdots$ for the other coin.
The second one is obtained by iterating $SX$ and results in the sequence $U_2 U_2 U_2 U_2\cdots$ for one coin and $U_1 U_1 U_1 U_1\cdots$ for the other. Both of them are analyzed numerically in~\cite{RajendranBenjamin2018}, and the first one is seen to be a winning game while the second one is a losing one (for appropriate choice of parameters for the coins and initial state).
The results are reported in Fig. 3--a) and Fig. 3--b) of~\cite{RajendranBenjamin2018}.

To repeat, we are not going to give a fully analytical proof of the numerical results in~\cite{RajendranBenjamin2018}, but rather we will use our results to explore the ``phase space'' of coins $U_1,U_2$ and separate the regions where the game obtained by iterating $SY SX$ is winning or losing.
This is a very good use of our analytical tools. 
The very last set of figures in this section explores the effect of slowly changing the initial state. In section $6$ we give numerical evidence that our two
games are a winning and a losing game respectively, in analogy to the games given in Fig. 3--a) and Fig. 3--b) of
\cite{RajendranBenjamin2018}.

We start with some useful observations.
Notice that $\sum_{x\in\mathbb{Z}}\mathbb{P}(X_t=x)=1$.
One can prove that $\Delta$, defined in Eq.~\eqref{eq:delta_Th2}, satisfies
\begin{equation}
 \Delta=\sum_{x\in\mathbb{Z}} \lim_{t\to\infty}\mathbb{P}(X_t=x) < 1,\label{eq:less_than_1}
\end{equation}
a clear indication that the summation and the limit can not be interchanged.
The inequality in Eq.~\eqref{eq:less_than_1} holds as long as we keep the limitation $\theta_1,\theta_2\notin\left\{0,\pi/2,\pi,3\pi/2\right\}$ which has been assumed since Sec.~\ref{sec:definition}.  
The reader may consult on this point section 1.2 of a book chapter, pages 337-350~\cite{Ko}, for the effect of localization.

A natural criterion to decide if a game is a winning or a losing one is to compare the quantities $P_R(t)$ and $P_L(t)$. 
They represent the probability $P_R(t)$ (resp. $P_L(t)$) that the walker is observed on the right (resp. left) side to the origin at time $t$,
\begin{align}
 P_R(t)=&\sum_{x=1}^{\infty} \mathbb{P}(X_t=x),\\
 P_L(t)=&\sum_{x=-\infty}^{-1} \mathbb{P}(X_t=x).
\end{align}
A careful analysis, along the lines used above, shows that this is given by the sum of two quantities that can be calculated using Theorems $1$ and $2$ above, namely
\begin{align}
 \mu_R=&\sum_{x=1}^\infty \lim_{t\to\infty}\mathbb{P}(X_t=x)\nonumber\\
 &+\int_0^\infty \frac{|s_m|\,(d_0+d_1x+d_2x^2)}{\,\pi(4-x^2)\sqrt{4c_m^2-x^2}\,}\,I_{\left(-2|c_m|,\,2|c_m|\right)}(x)\,dx,\label{eq:mu_R}\\
 \mu_L=&\sum_{x=-\infty}^{-1} \lim_{t\to\infty}\mathbb{P}(X_t=x)\nonumber\\
 &+\int_{-\infty}^0 \frac{|s_m|\,(d_0+d_1x+d_2x^2)}{\,\pi(4-x^2)\sqrt{4c_m^2-x^2}\,}\,I_{\left(-2|c_m|,\,2|c_m|\right)}(x)\,dx.\label{eq:mu_L}
\end{align}
The first summand in each case is very hard to obtain analytically so we choose a large value of $n$ and replace $\mu_R, \mu_L$ by
\begin{align}
 \mu_R(n)=&\sum_{x=1}^n \lim_{t\to\infty}\mathbb{P}(X_t=x)\nonumber\\
 &+\int_0^\infty \frac{|s_m|\,(d_0+d_1x+d_2x^2)}{\,\pi(4-x^2)\sqrt{4c_m^2-x^2}\,}\,I_{\left(-2|c_m|,\,2|c_m|\right)}(x)\,dx,\label{eq:mu_R(n)}\\
 \mu_L(n)=&\sum_{x=-n}^{-1} \lim_{t\to\infty}\mathbb{P}(X_t=x)\nonumber\\
 &+\int_{-\infty}^0 \frac{|s_m|\,(d_0+d_1x+d_2x^2)}{\,\pi(4-x^2)\sqrt{4c_m^2-x^2}\,}\,I_{\left(-2|c_m|,\,2|c_m|\right)}(x)\,dx.\label{eq:mu_L(n)}
\end{align}
respectively.
Note that the value of $\lim_{t\to\infty}\mathbb{P}(X_t=x)$ exponentially decays as the location $x$ is going far away from the origin $x=0$.
Ignoring such small values in Eqs.~\eqref{eq:mu_R} and \eqref{eq:mu_L} results in Eqs.~\eqref{eq:mu_R(n)} and \eqref{eq:mu_L(n)} with a large value of $n$.
With these analytical tools we start exploring the parameter space of the 2-period time-dependent quantum walk studied here.

\clearpage

\begin{figure}[h]
\begin{center}
 \begin{minipage}{50mm}
  \begin{center}
  \includegraphics[scale=0.4]{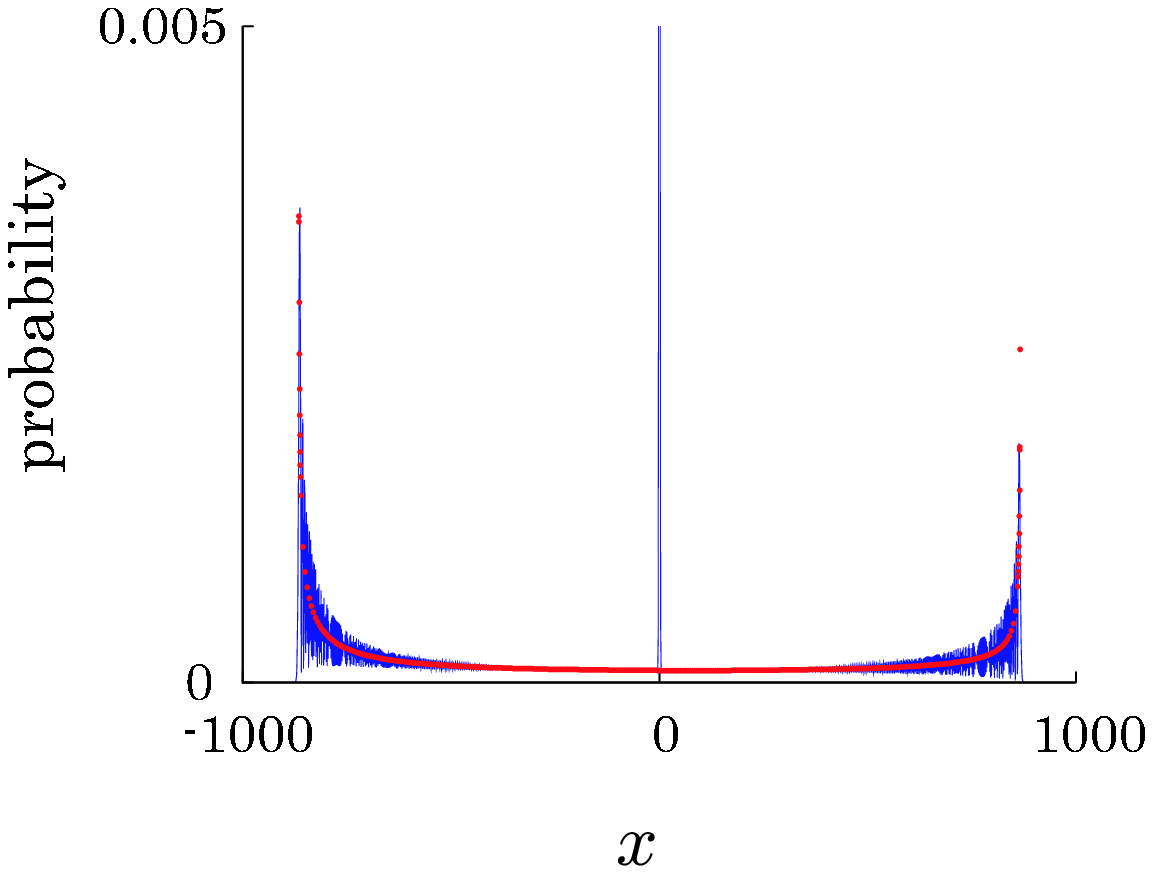}\\[2mm]
  (a) $\theta_1=\pi/6, \theta_2=\pi/6$
  \end{center}
 \end{minipage}
 \begin{minipage}{50mm}
  \begin{center}
 \includegraphics[scale=0.4]{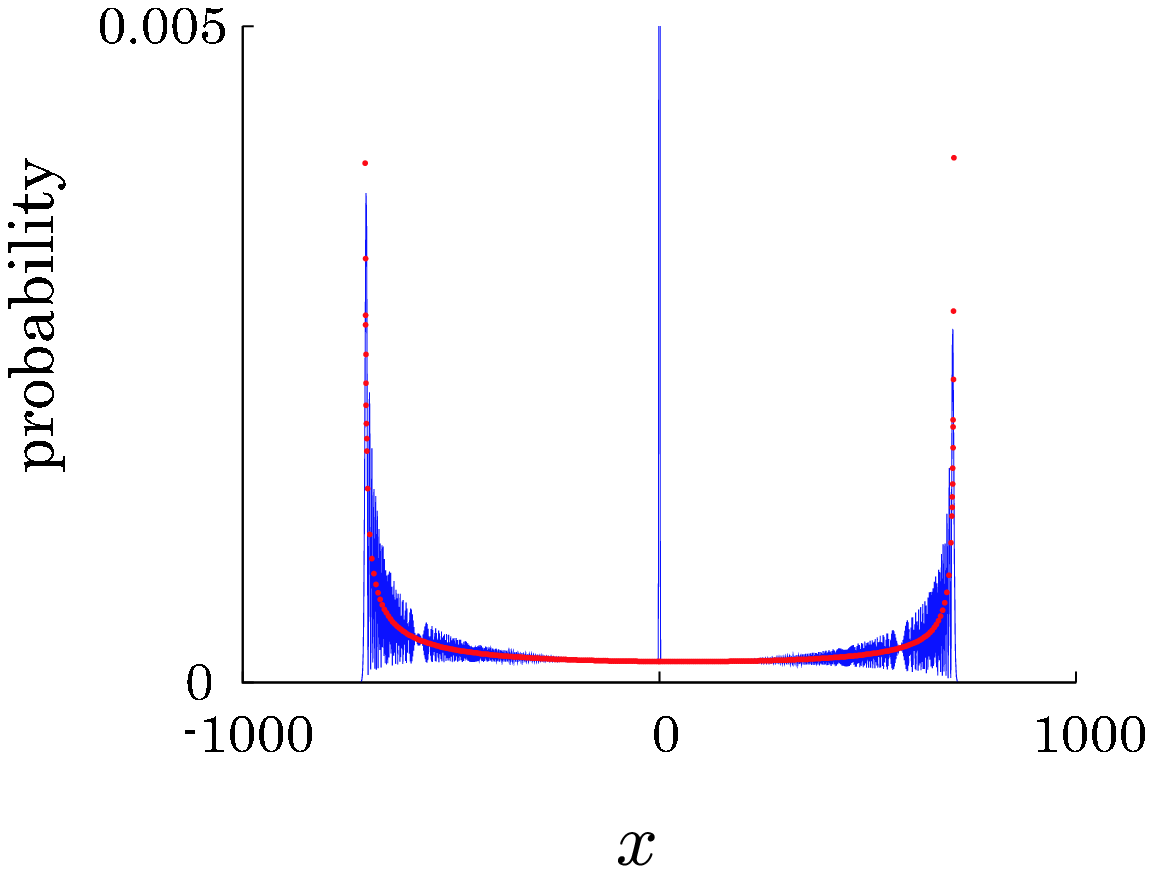}\\[2mm]
  (b) $\theta_1=\pi/4, \theta_2=\pi/4$
  \end{center}
 \end{minipage}
 \vspace{5mm}

 \begin{minipage}{50mm}
  \begin{center}
  \includegraphics[scale=0.4]{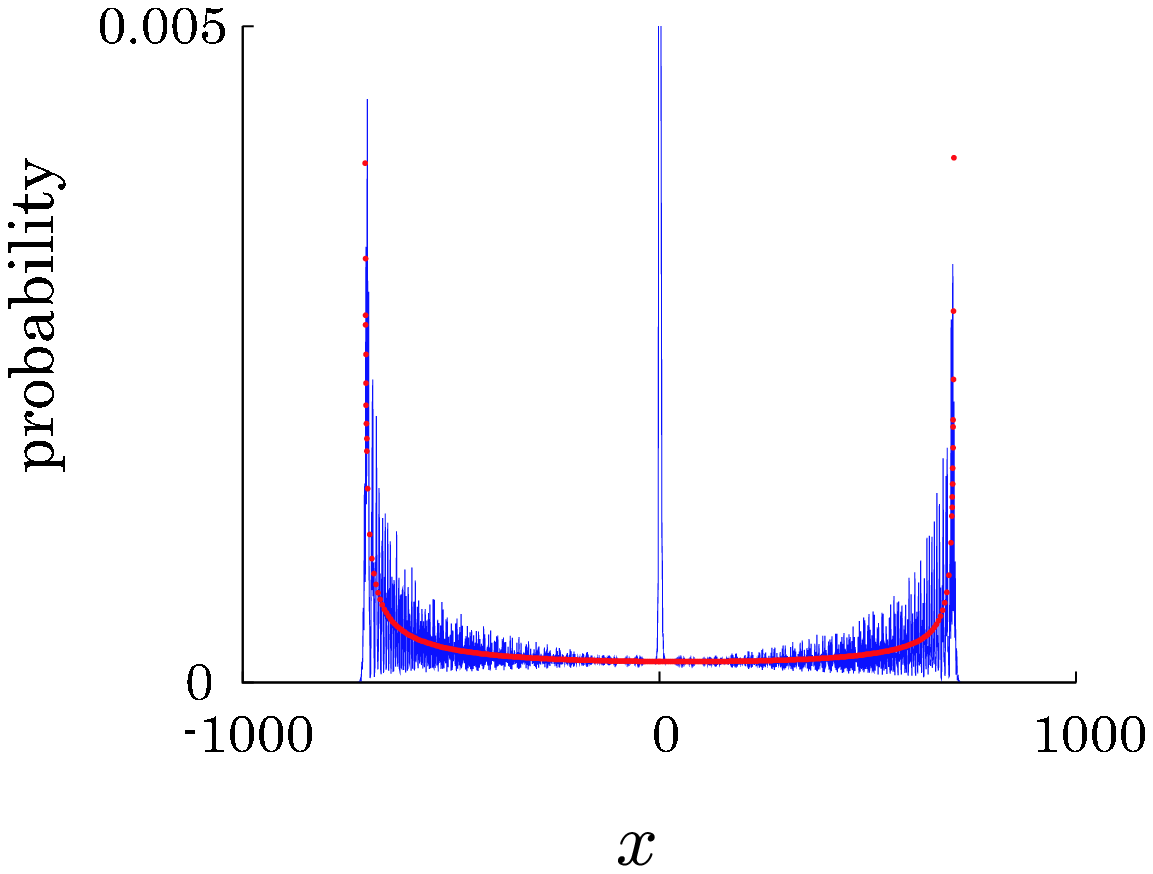}\\[2mm]
  (c) $\theta_1=\pi/6, \theta_2=\pi/4$
  \end{center}
 \end{minipage}
 \begin{minipage}{50mm}
  \begin{center}
 \includegraphics[scale=0.4]{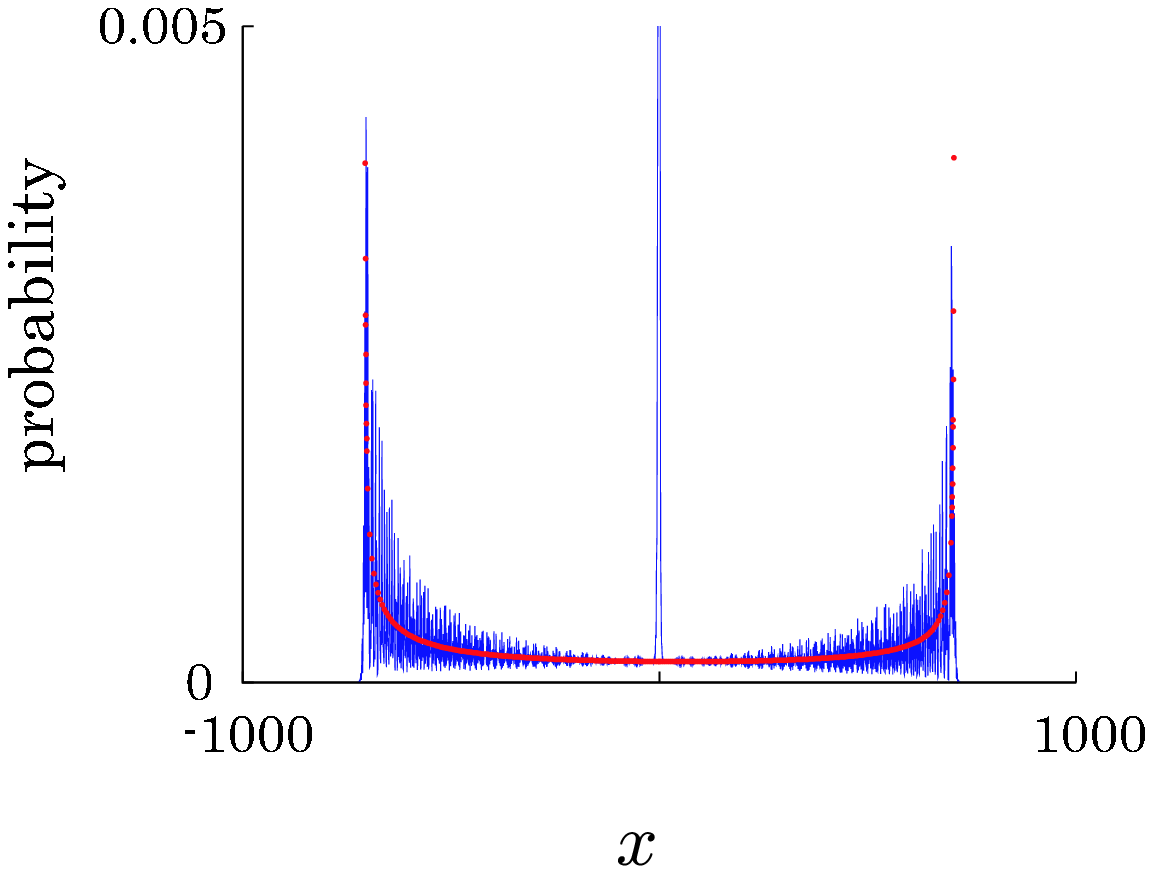}\\[2mm]
  (d) $\theta_1=\pi/4, \theta_2=\pi/6$
  \end{center}
 \end{minipage}
 \vspace{5mm}
 \caption{$\mathbb{P}(X_{500}=x)$ (blue line) and an approximation obtained from the continuous-part of the limit density function (red points) with $q_0=\cos(5\pi/12), q_1=i\sin(5\pi/12), q_2=0, q_3=0$}
 \label{fig:fig13}
\end{center}
\end{figure}

\begin{figure}[h]
\begin{center}
 \begin{minipage}{50mm}
  \begin{center}
  \includegraphics[scale=0.4]{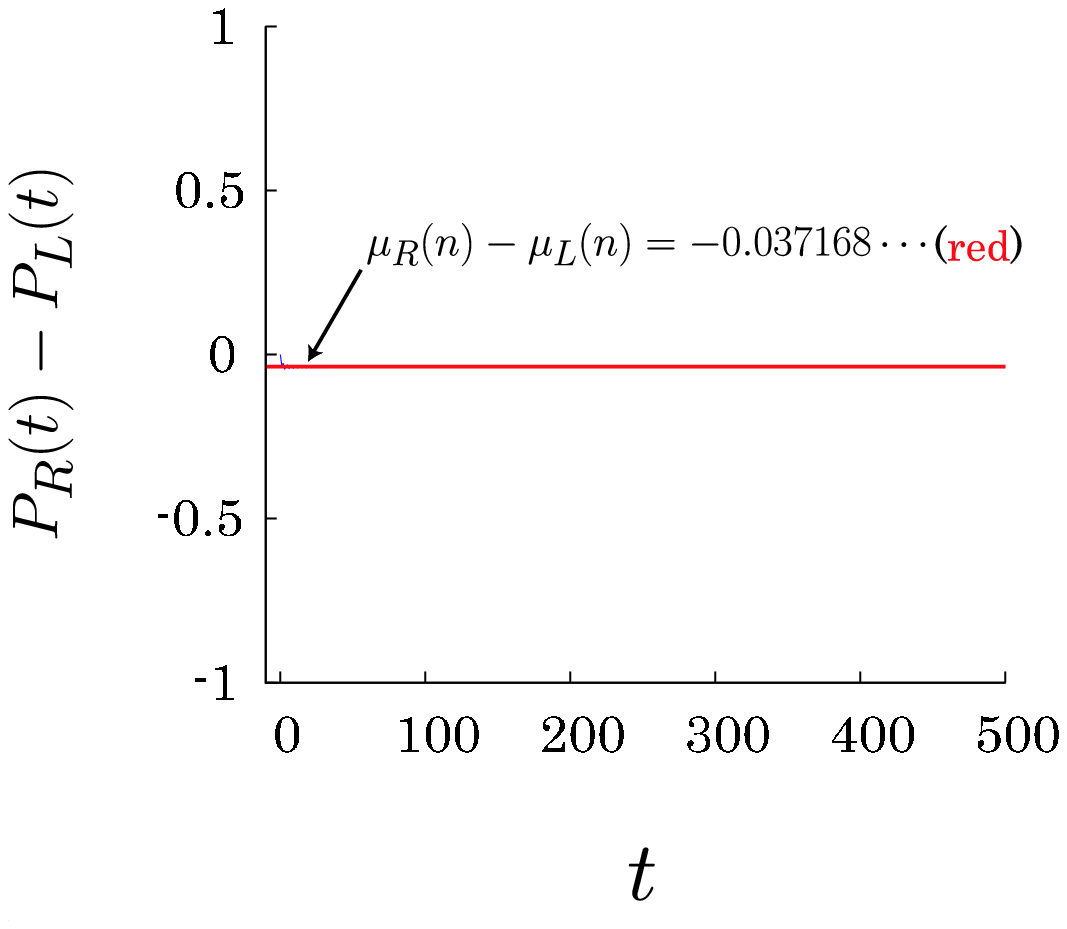}\\[2mm]
  (a) $\theta_1=\pi/6, \theta_2=\pi/6$
  \end{center}
 \end{minipage}
 \begin{minipage}{50mm}
  \begin{center}
 \includegraphics[scale=0.4]{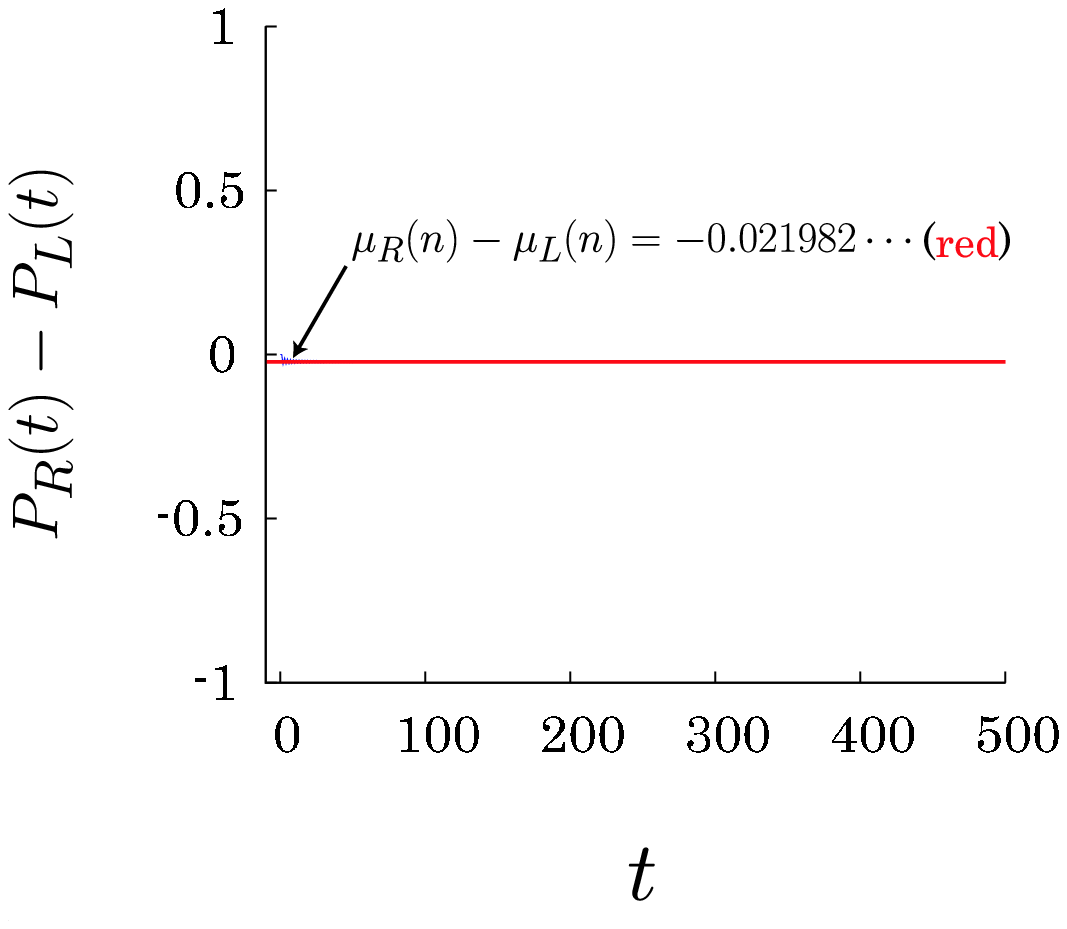}\\[2mm]
  (b) $\theta_1=\pi/4, \theta_2=\pi/4$
  \end{center}
 \end{minipage}
 \vspace{5mm}

 \begin{minipage}{50mm}
  \begin{center}
  \includegraphics[scale=0.4]{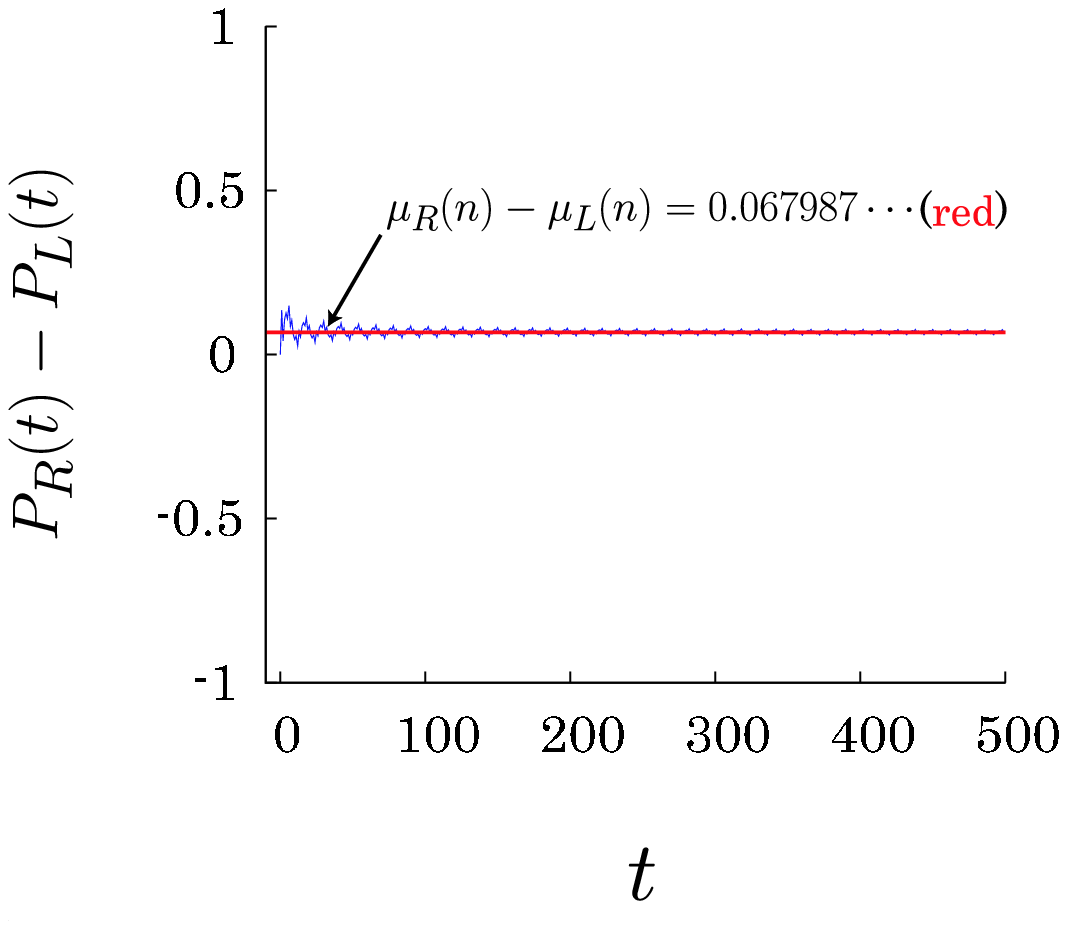}\\[2mm]
  (c) $\theta_1=\pi/6, \theta_2=\pi/4$
  \end{center}
 \end{minipage}
 \begin{minipage}{50mm}
  \begin{center}
 \includegraphics[scale=0.4]{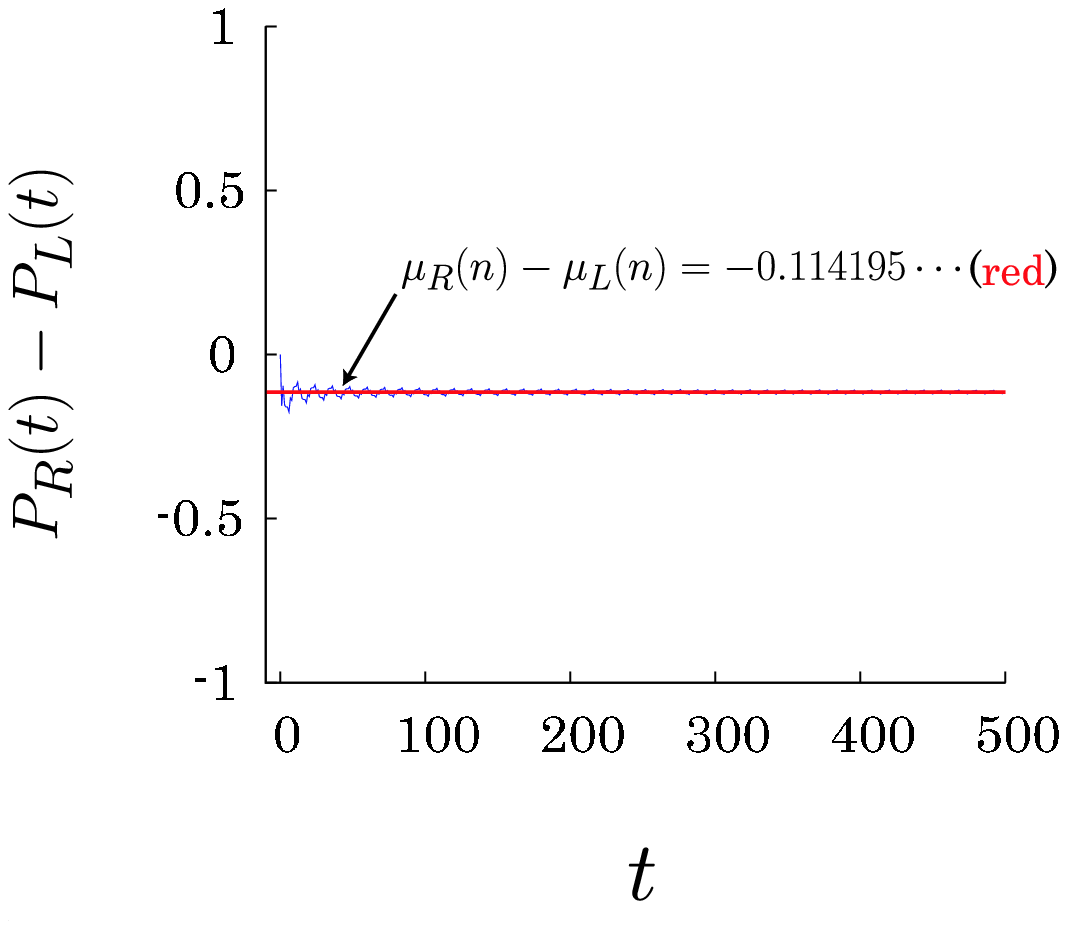}\\[2mm]
  (d) $\theta_1=\pi/4, \theta_2=\pi/6$
  \end{center}
 \end{minipage}
 \vspace{5mm}
 \caption{$P_R(t)-P_L(t)$ (blue line) and $\mu_R(n)-\mu_L(n)\,\,(n=10000)$ by approximately numerical experiment (red line) with $q_0=\cos(5\pi/12), q_1=i\sin(5\pi/12), q_2=0, q_3=0$ : Only the case of parameters $\theta_1=\pi/6, \theta_2=\pi/4$ results in the winning game ((c)) and the others the losing games ((a), (b), (d)).}
 \label{fig:fig12}
\end{center}
\end{figure}

Fixing the initial state at
\begin{equation}
 \ket{\Psi_0}=\ket{0}\otimes\left(\cos\left(\frac{5\pi}{12}\right)\ket{00}+i\sin\left(\frac{5\pi}{12}\right)\ket{01}\right),
\end{equation}
we see the probability distribution at time $500$ and the limit distribution in Figure~\ref{fig:fig13}, or $P_R(t)-P_L(t)$ and $\mu_R(n)-\mu_L(n)\,(n=10000)$ in Fig.~\ref{fig:fig12}. 

\clearpage

\clearpage

\begin{figure}[h]
\begin{center}
 \begin{minipage}{100mm}
  \begin{center}
  \includegraphics[scale=0.7]{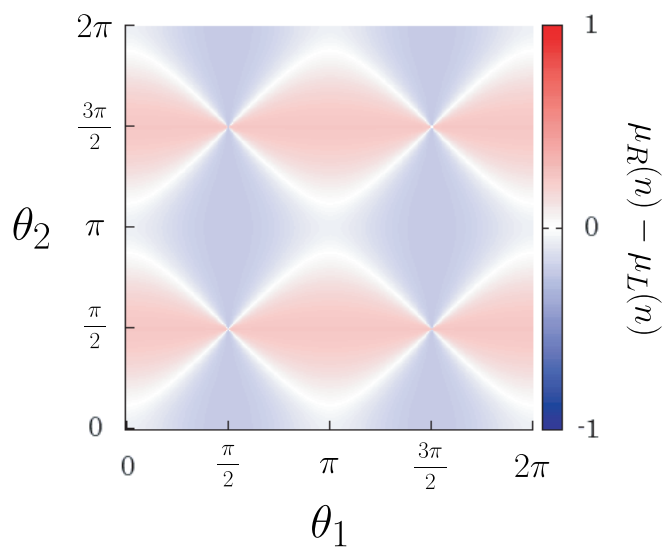}\\[2mm]
  (a) $q_0=\cos(5\pi/12), q_1=i\sin(5\pi/12), q_2=0, q_3=0$
  \end{center}
 \end{minipage}
 \vspace{5mm}
 \begin{minipage}{100mm}
  \begin{center}
 \includegraphics[scale=0.7]{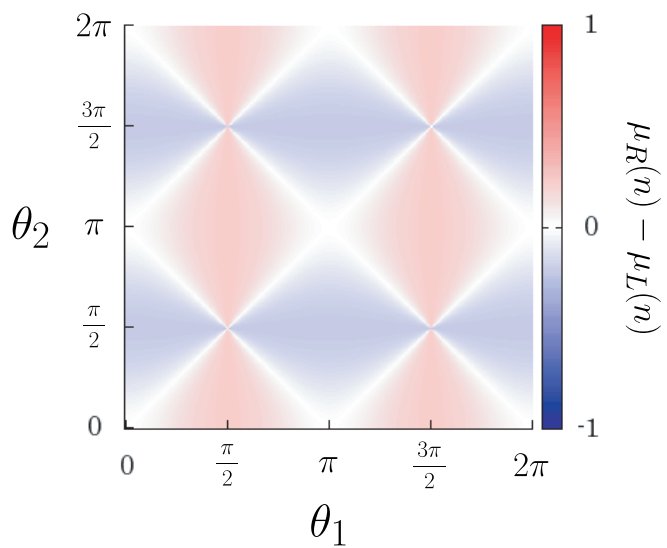}\\[2mm]
  (b) $q_0=0, q_1=\cos(5\pi/12), q_2=i\sin(5\pi/12), q_3=0$
  \end{center}
 \end{minipage}
 \vspace{5mm}
 \caption{$\mu_R(n)-\mu_L(n)\,\,(n=10000)$ : The pair of parameters $\theta_1$ and $\theta_2$ in the red colored area gives a winning game, in the blue colored area gives a losing one, and the white colored area corresponds to a draw.}
 \label{fig:fig14-ab}
 \end{center}
\end{figure}

\begin{figure}[h]
\begin{center}
 \begin{minipage}{100mm}
  \begin{center}
  \includegraphics[scale=0.7]{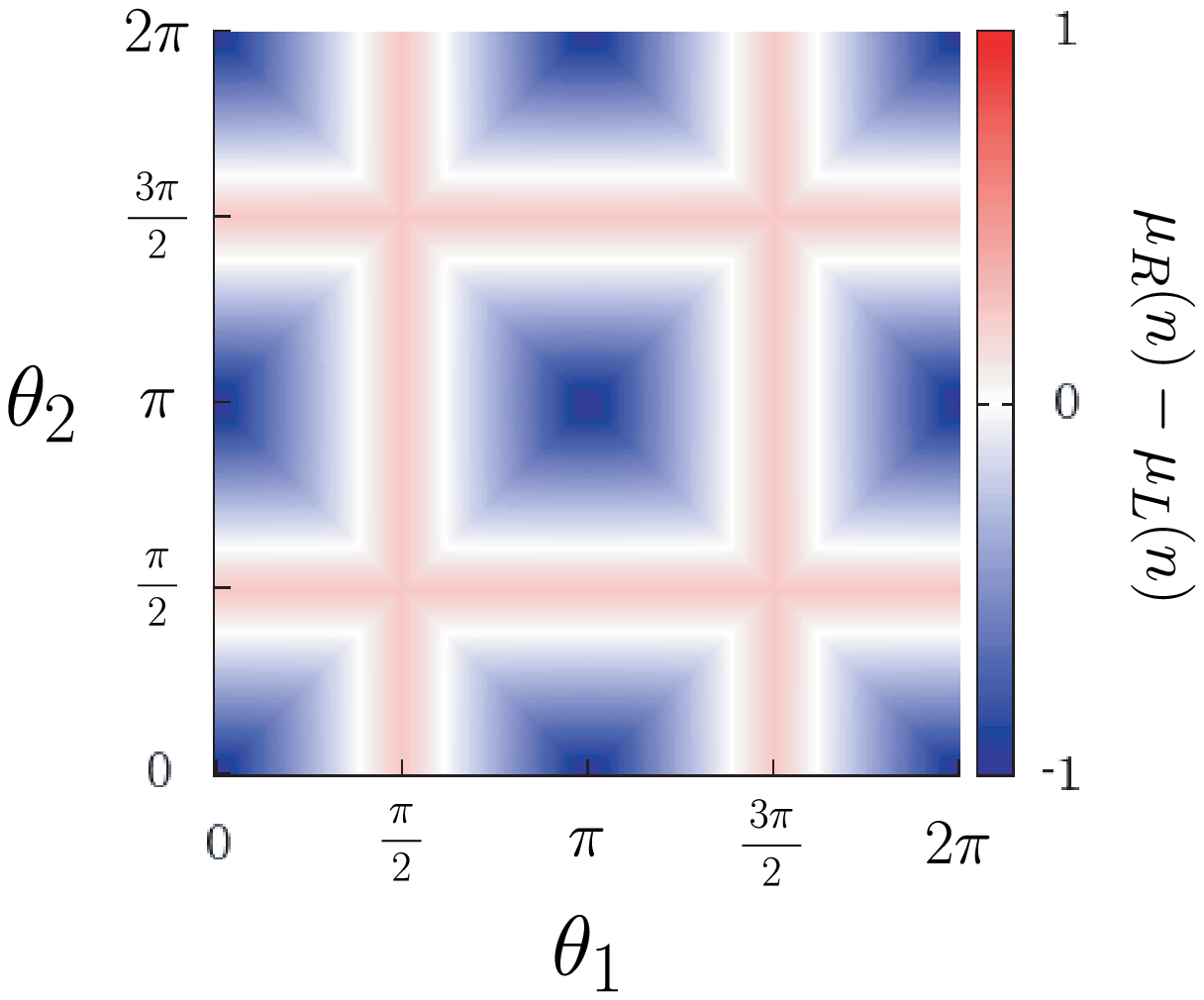}\\[2mm]
  (c) $q_0=1, q_1=0, q_2=0, q_3=0$
  \end{center}
 \end{minipage}
 \vspace{5mm}
 \begin{minipage}{100mm}
  \begin{center}
 \includegraphics[scale=0.7]{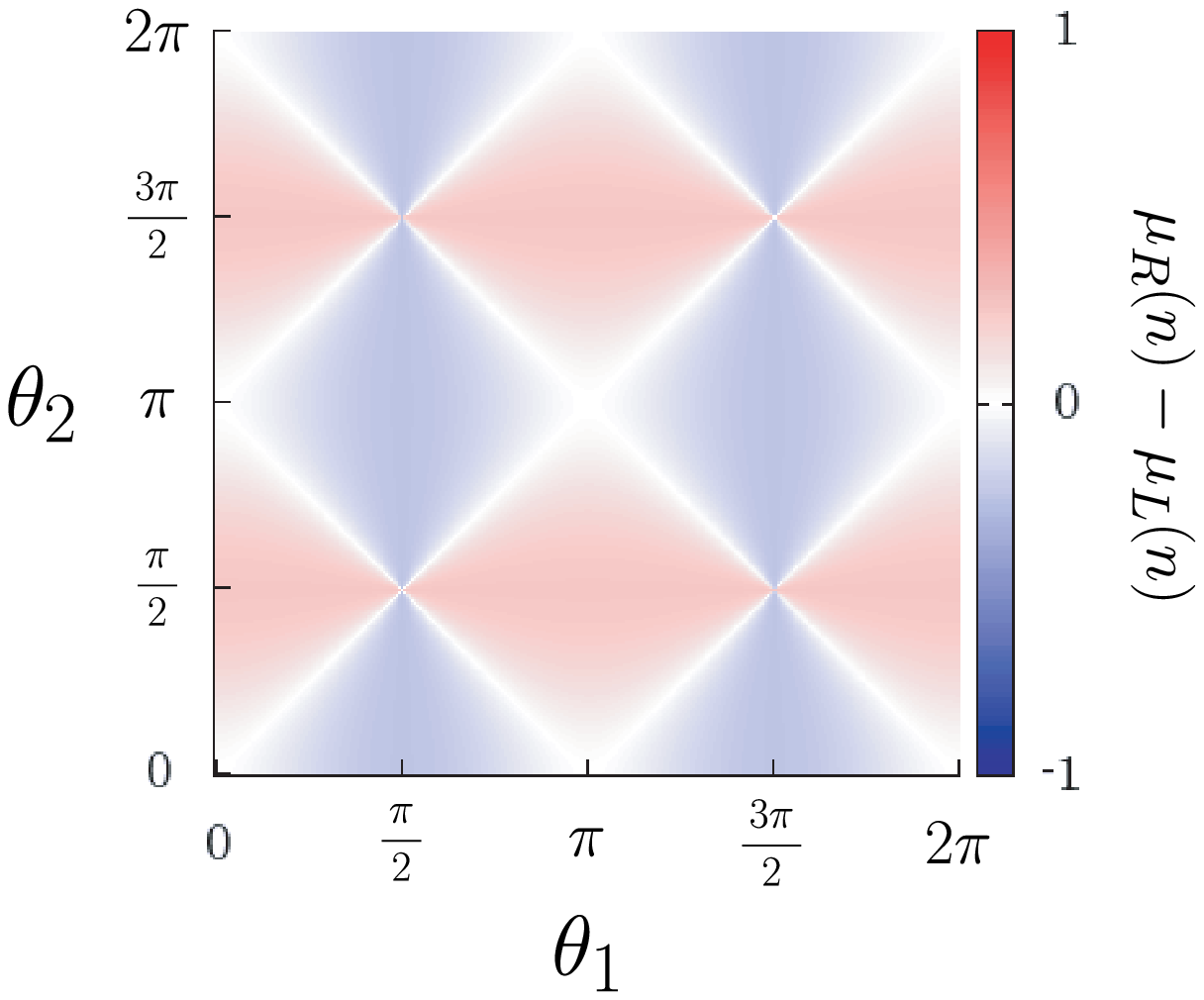}\\[2mm]
  (d) $q_0=0, q_1=1, q_2=0, q_3=0$
  \end{center}
 \end{minipage}
 \vspace{5mm}
 \caption{$\mu_R(n)-\mu_L(n)\,\,(n=10000)$ : The pair of parameters $\theta_1$ and $\theta_2$ in the red colored area corresponds to a winning game, in the blue colored area to a losing one, and the white colored corresponds to a draw.}
 \label{fig:fig14-cd}
 \end{center}
\end{figure}

\begin{figure}[h]
\begin{center}
 \begin{minipage}{100mm}
  \begin{center}
  \includegraphics[scale=0.7]{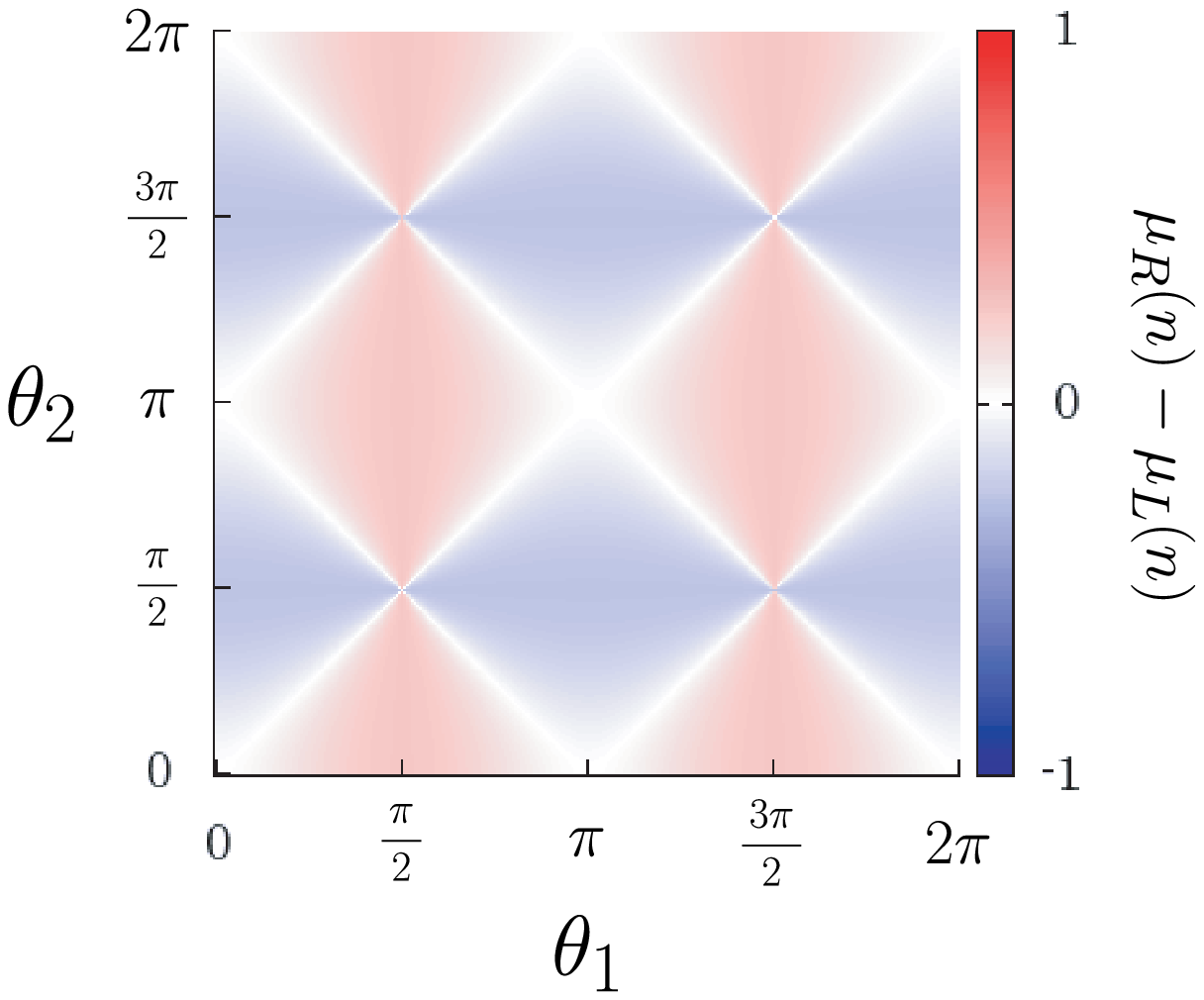}\\[2mm]
  (e) $q_0=0, q_1=0, q_2=1, q_3=0$
  \end{center}
 \end{minipage}
 \vspace{5mm}
 \begin{minipage}{100mm}
  \begin{center}
 \includegraphics[scale=0.7]{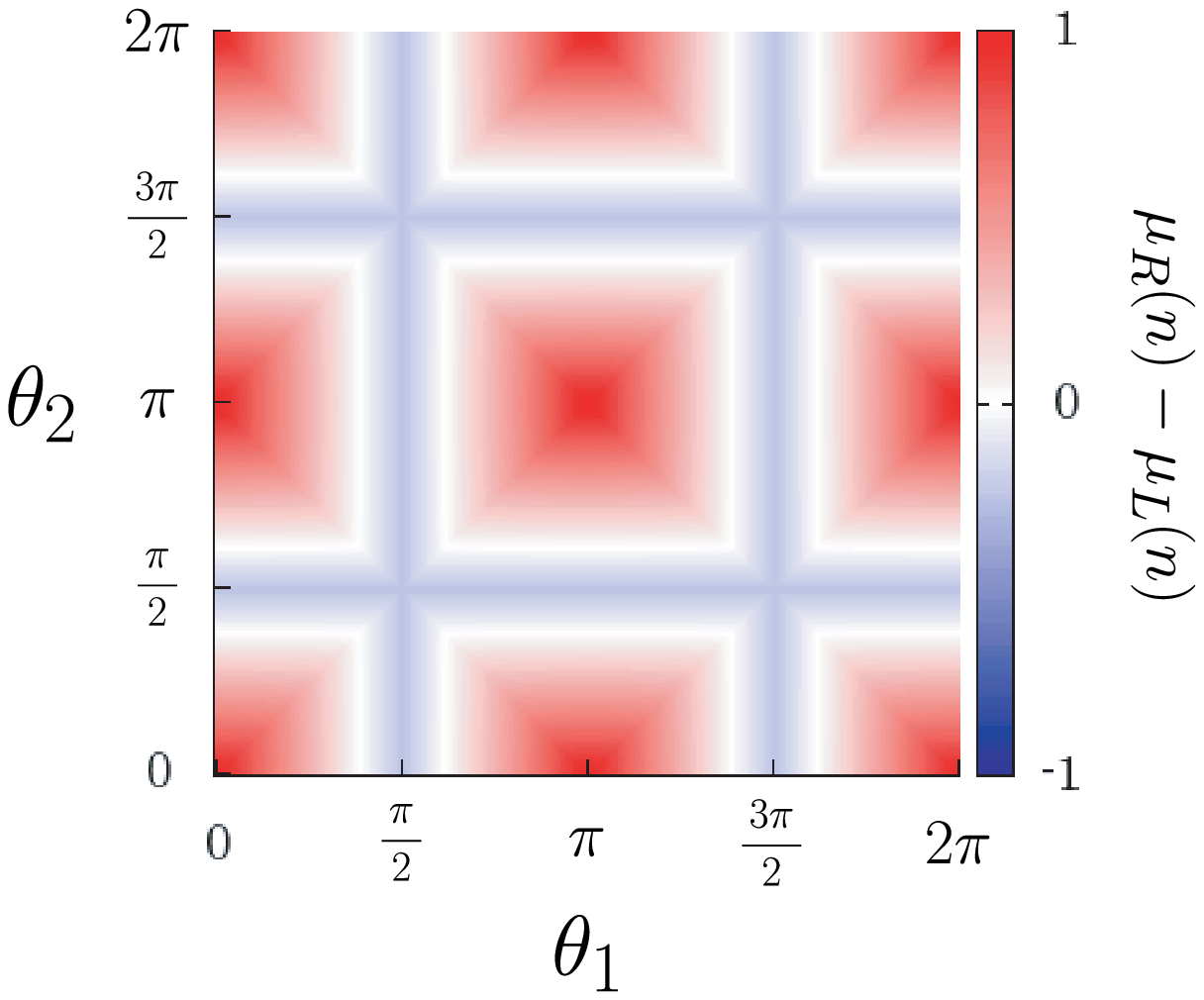}\\[2mm]
  (f) $q_0=0, q_1=0, q_2=0, q_3=1$
  \end{center}
 \end{minipage}
 \vspace{5mm}
 \caption{$\mu_R(n)-\mu_L(n)\,\,(n=10000)$ : The pair of parameters $\theta_1$ and $\theta_2$ in the red colored area corresponds to a winning game, in the blue colored area to a losing one, and the white colored corresponds to a draw.}
 \label{fig:fig14-ef}
 \end{center}
\end{figure}

\begin{figure}[h]
\begin{center}
 \begin{minipage}{100mm}
  \begin{center}
  \includegraphics[scale=0.7]{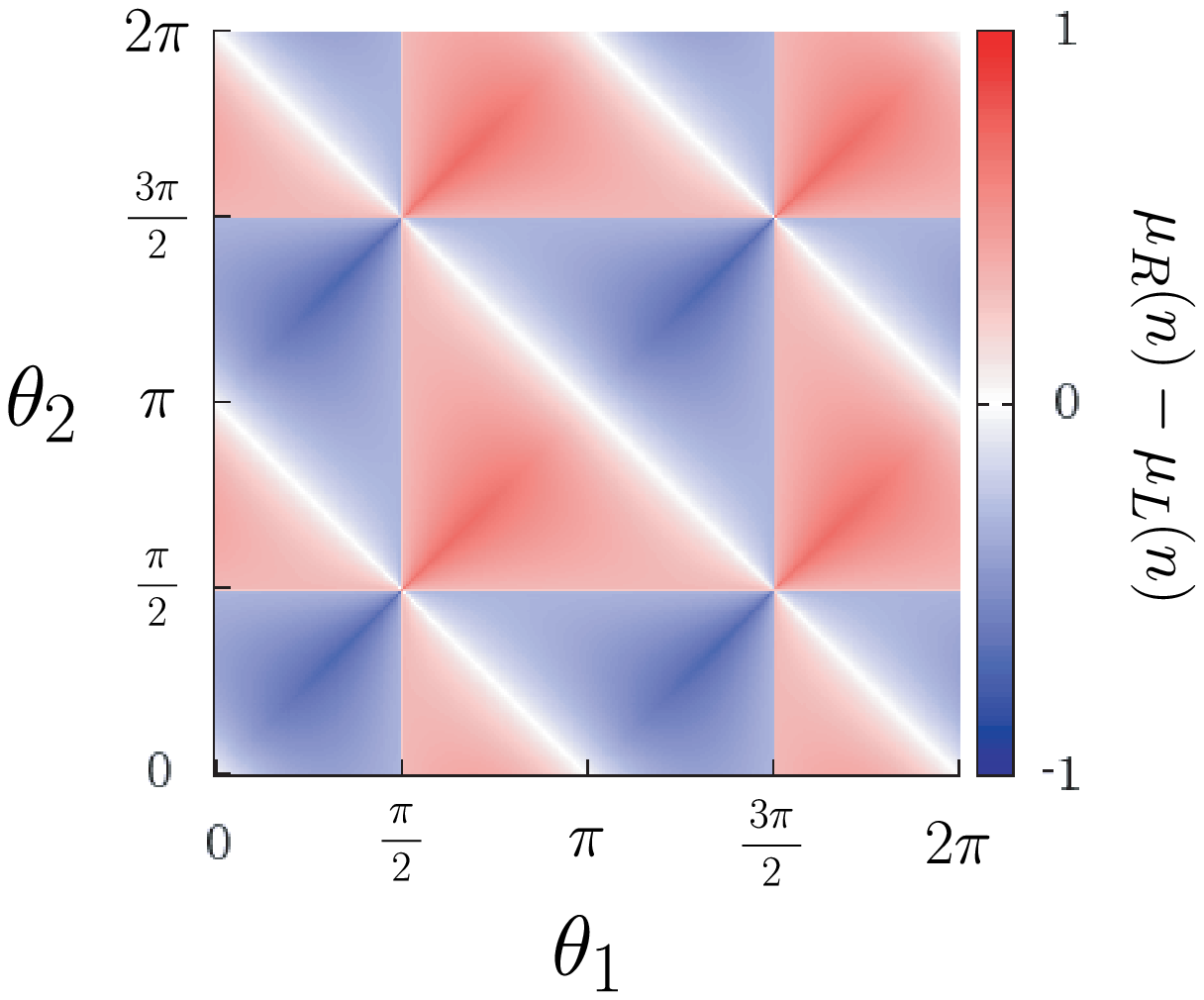}\\[2mm]
  (g) $q_0=1/2, q_1=1/2, q_2=1/2, q_3=1/2$
  \end{center}
 \end{minipage}
 \vspace{5mm}
 \caption{$\mu_R(n)-\mu_L(n)\,\,(n=10000)$ : The pair of parameters $\theta_1$ and $\theta_2$ in the red colored area corresponds to a winning game, in the blue colored area to a losing one, and the white colored corresponds to a draw.}
 \label{fig:fig14-g}
 \end{center}
\end{figure}

Figures~\ref{fig:fig14-ab}--\ref{fig:fig14-g} display the values of $\mu_R(n)-\mu_L(n)\,\,(n=10000)$ and report what values of the parameters $\theta_1$ and $\theta_2$ give a winning or a losing game.
For the initial state indicated in each caption, we find the regions where we have a winning or a losing game.

\clearpage

\begin{figure}[h]
\begin{center}
 \begin{minipage}{50mm}
  \begin{center}
  \includegraphics[scale=0.4]{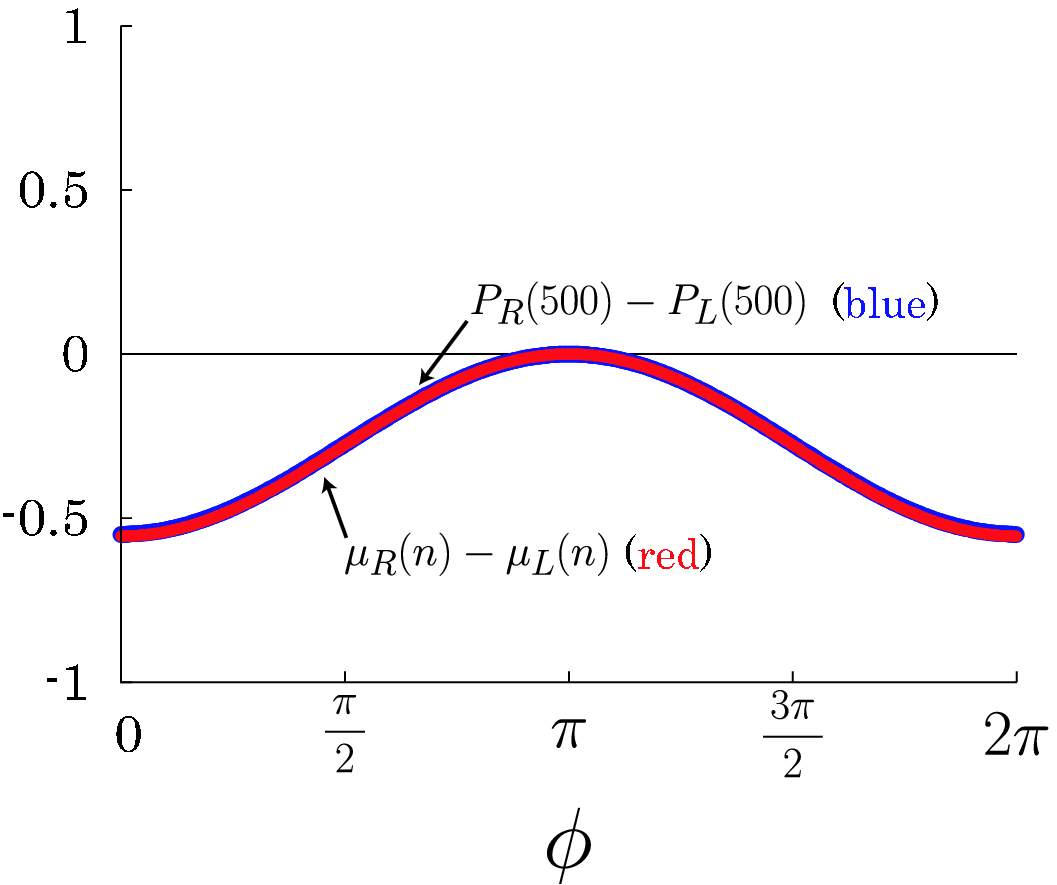}\\[2mm]
  (a) $\theta_1=\pi/6, \theta_2=\pi/6$
  \end{center}
 \end{minipage}
 \begin{minipage}{50mm}
  \begin{center}
 \includegraphics[scale=0.4]{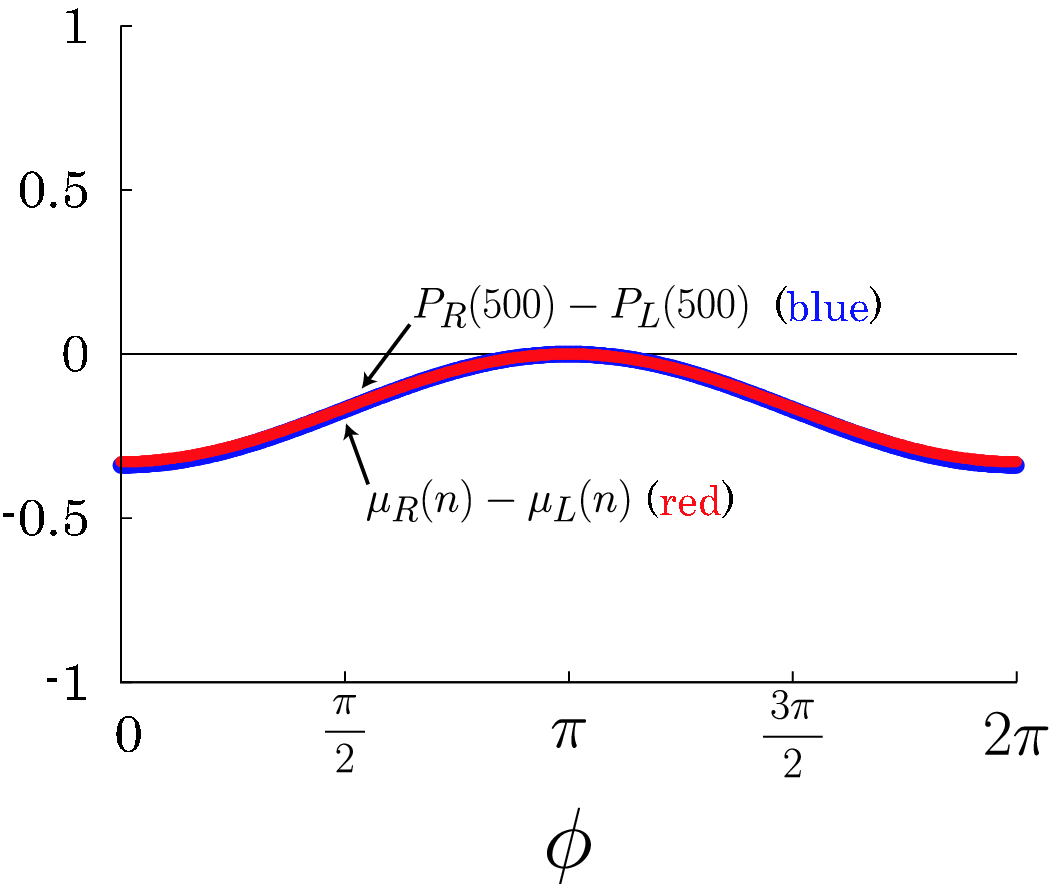}\\[2mm]
  (b) $\theta_1=\pi/4, \theta_2=\pi/4$
  \end{center}
 \end{minipage}
 \vspace{5mm}

 \begin{minipage}{50mm}
  \begin{center}
  \includegraphics[scale=0.4]{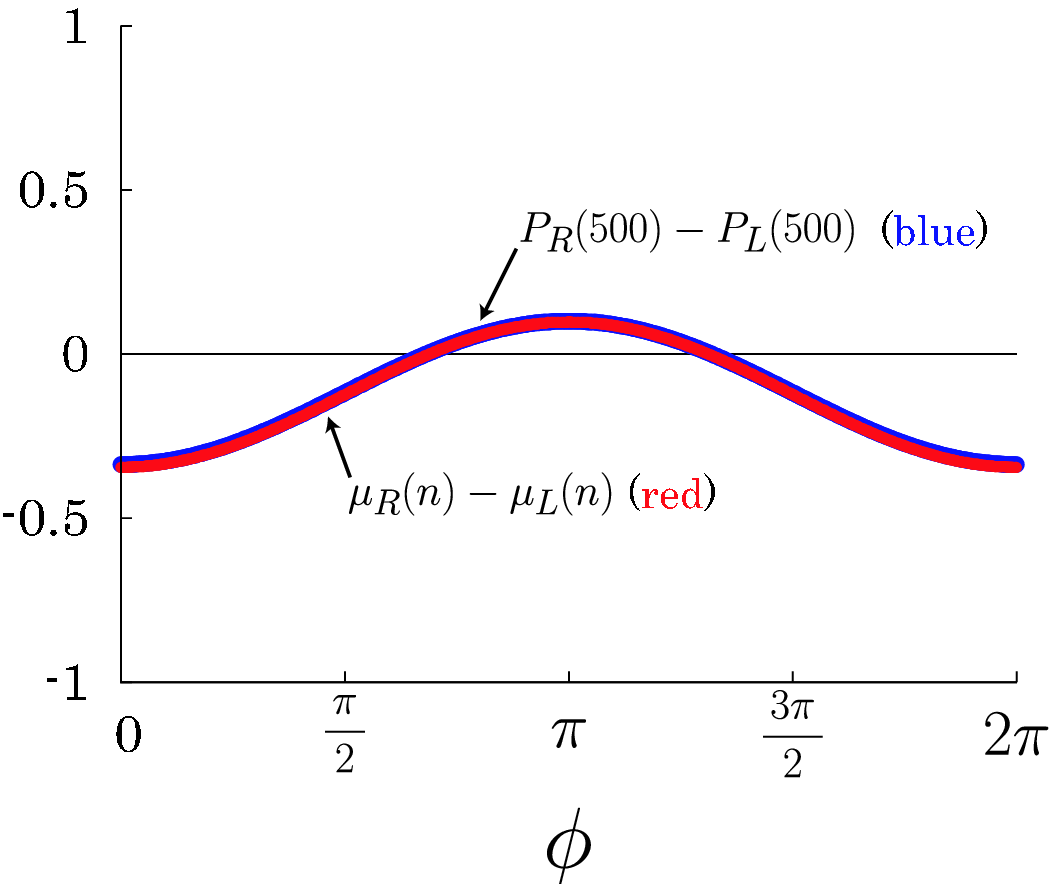}\\[2mm]
  (c) $\theta_1=\pi/6, \theta_2=\pi/4$
  \end{center}
 \end{minipage}
 \begin{minipage}{50mm}
  \begin{center}
 \includegraphics[scale=0.4]{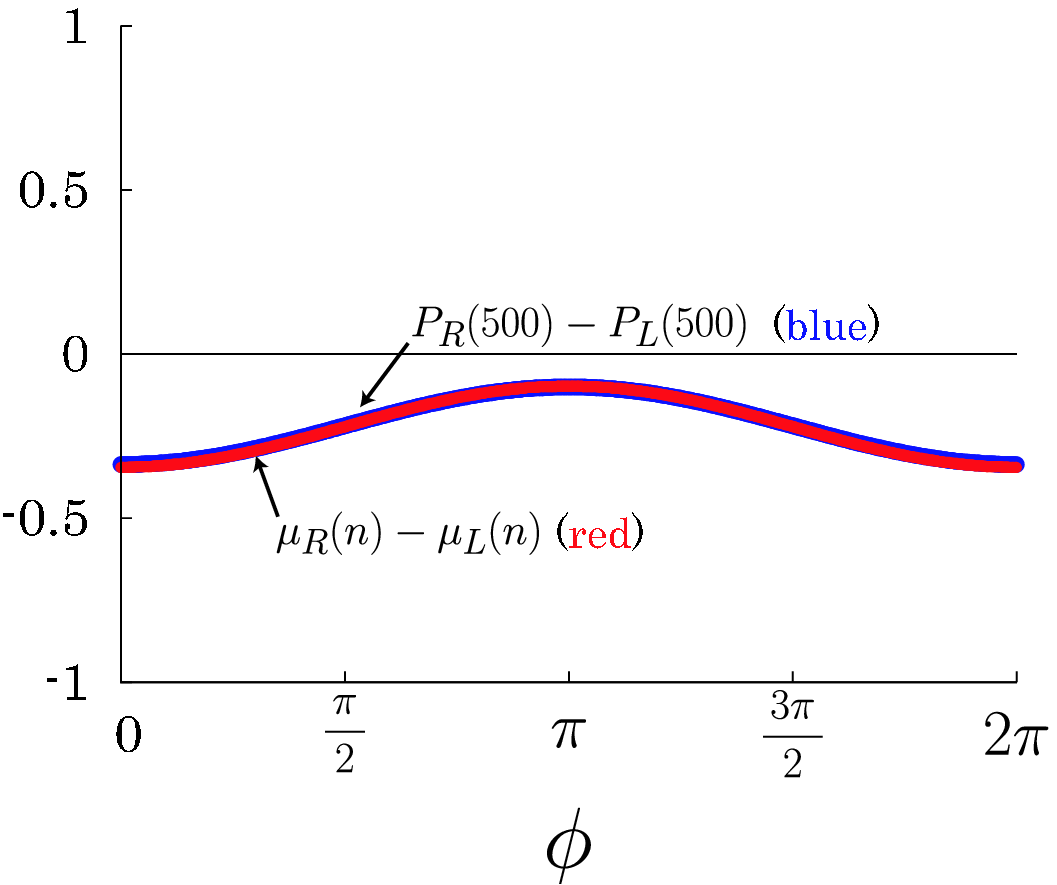}\\[2mm]
  (d) $\theta_1=\pi/4, \theta_2=\pi/6$
  \end{center}
 \end{minipage}
 \vspace{5mm}
	\caption{$P_R(500)-P_L(500)$ (blue points) and $\mu_R(n)-\mu_L(n)\,\,(n=10000)$ by approximately numerical experiment (red points) with $q_0=\cos(\phi/2), q_1=i\sin(\phi/2), q_2=0, q_3=0$ : The pair of parameters $\theta_1$ and $\theta_2$ used in each case make(c) a winning game, (d) a losing one, while (a) and (b) are losing ones except for a draw for one values of $\phi$.}
 \label{fig:fig11}
\end{center}
\end{figure}

This last set of figures shows the effect of varying the initial state on four different situations.
Again all these results are computed using our analytical formulas. 
Given the initial state in the form
\begin{equation}
 \ket{\Psi_0}=\ket{0}\otimes\left(\cos\left(\frac{\phi}{2}\right)\ket{00}+i\sin\left(\frac{\phi}{2}\right)\ket{01}\right),
\end{equation}
our results match the values obtained from $P_R(t)-P_L(t)$ at a large time $t=500$, as shown in Fig.~\ref{fig:fig11}.

\section{A numerical comparison of our 2-period time-dependent quantum walk with a time independent quantum walk}

Recall that we have studied analytically the game obtained by iterating
\begin{equation}
 SY SX,
\end{equation}
which results in the sequence $U_1 U_2 U_1 U_2\cdots$ for one coin and the sequence $U_2 U_1 U_2 U_1\cdots$ for the other coin.

\begin{figure}[h]
\begin{center}
 \begin{minipage}{50mm}
  \begin{center}
  \includegraphics[scale=0.4]{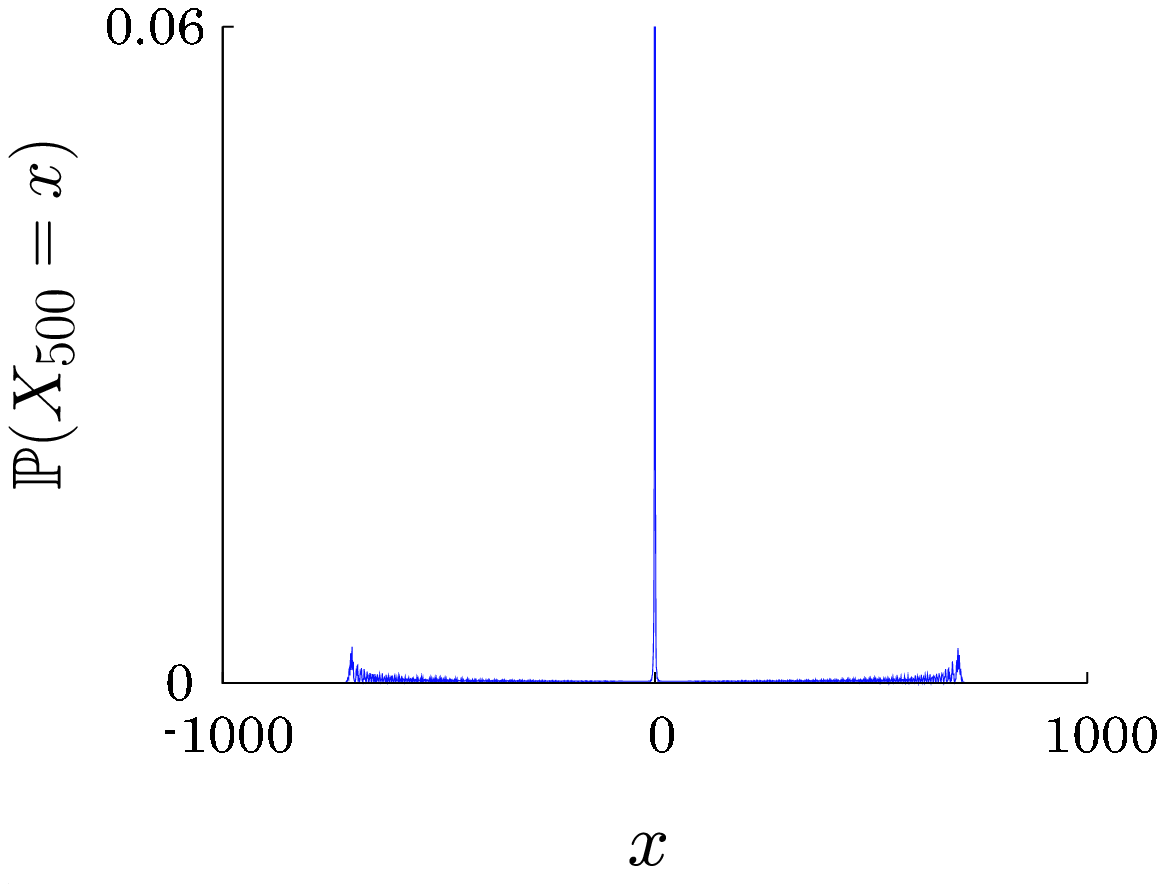}\\[2mm]
  \end{center}
 \end{minipage}
 \begin{minipage}{50mm}
  \begin{center}
 \includegraphics[scale=0.4]{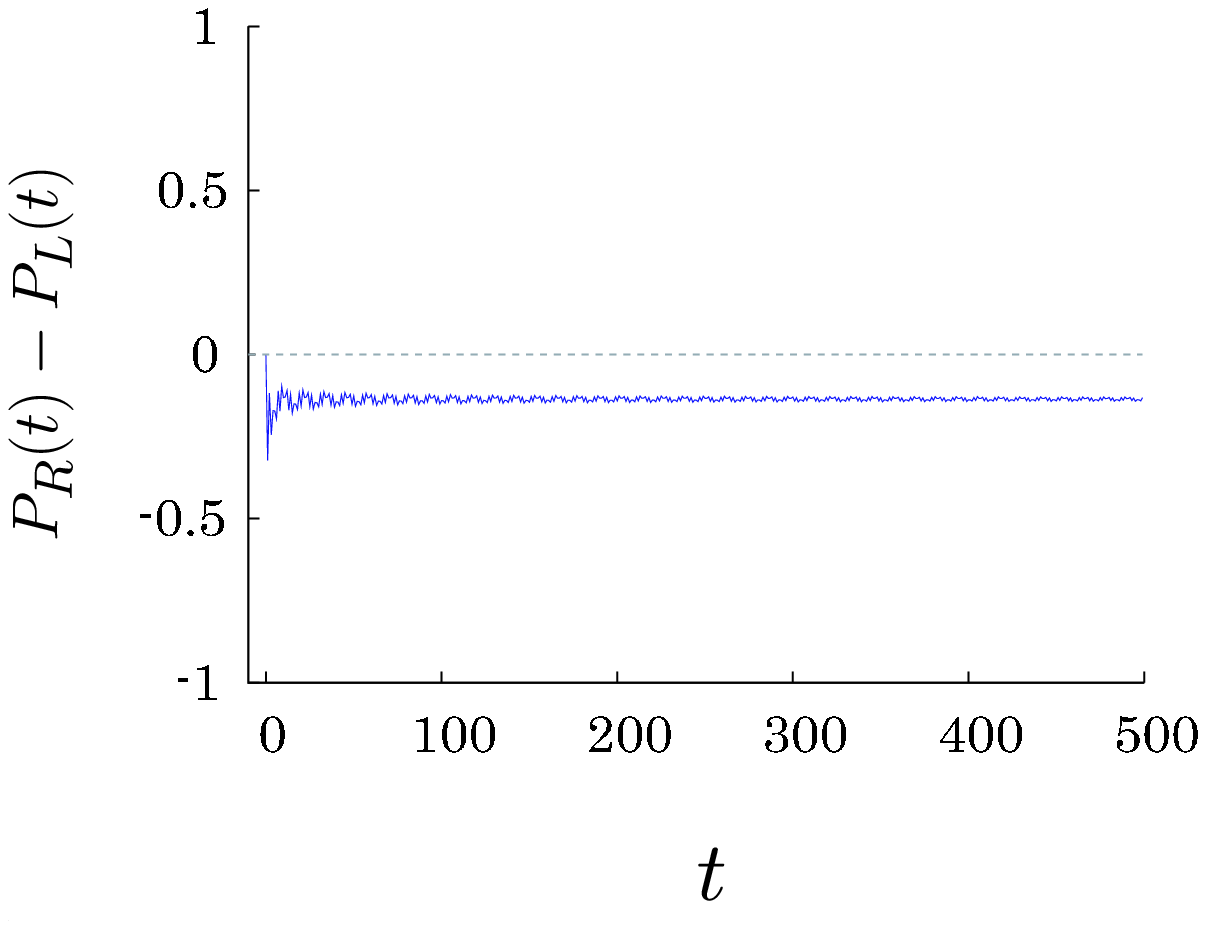}\\[2mm]
  \end{center}
 \end{minipage}
  \bigskip

 (a) $\ket{\Psi_t}=(SYSX)^t\Bigl\{\ket{0}\otimes 1/\sqrt{3}\,\Bigl(\ket{01}+i\ket{10}+\ket{11}\Bigr)\Bigr\}$ with $\theta_1=\pi/6, \theta_2=\pi/4$
 \vspace{5mm}

 \begin{minipage}{50mm}
  \begin{center}
  \includegraphics[scale=0.4]{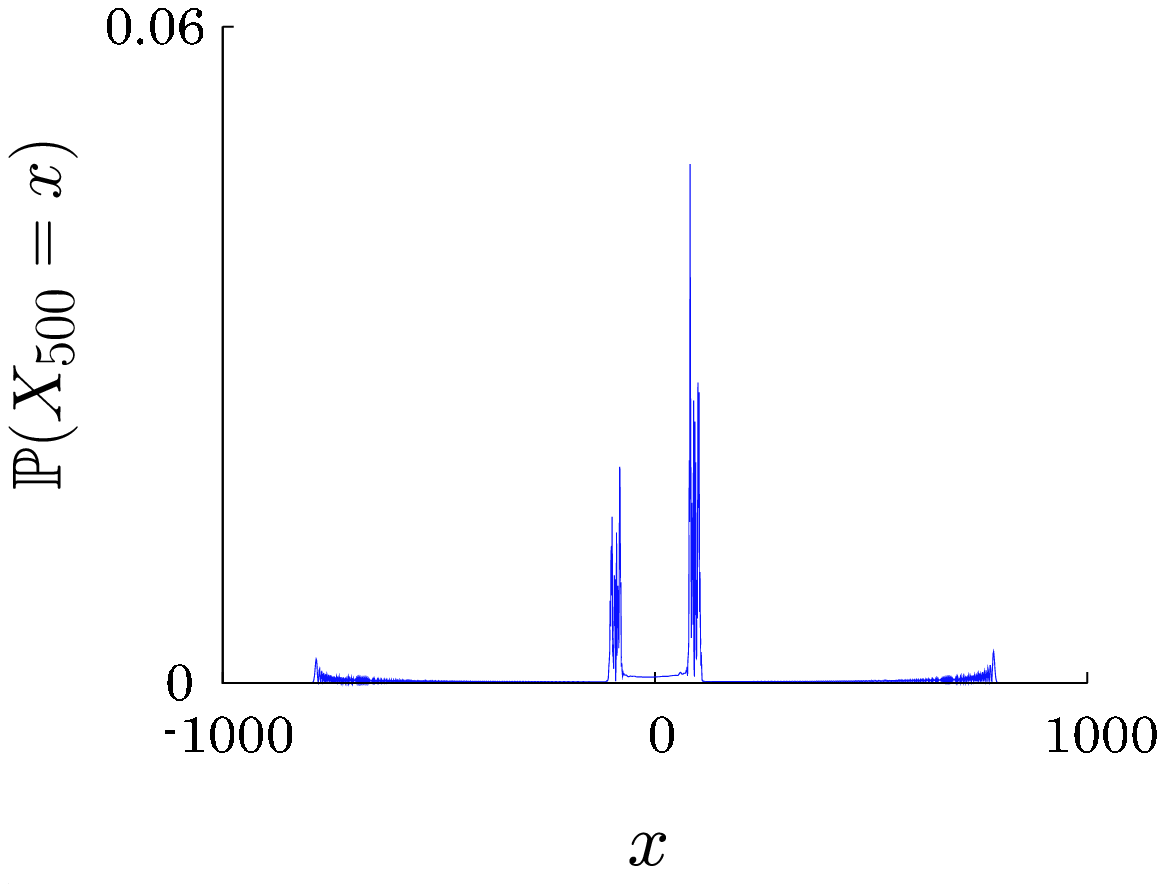}\\[2mm]
  \end{center}
 \end{minipage}
 \begin{minipage}{50mm}
  \begin{center}
 \includegraphics[scale=0.4]{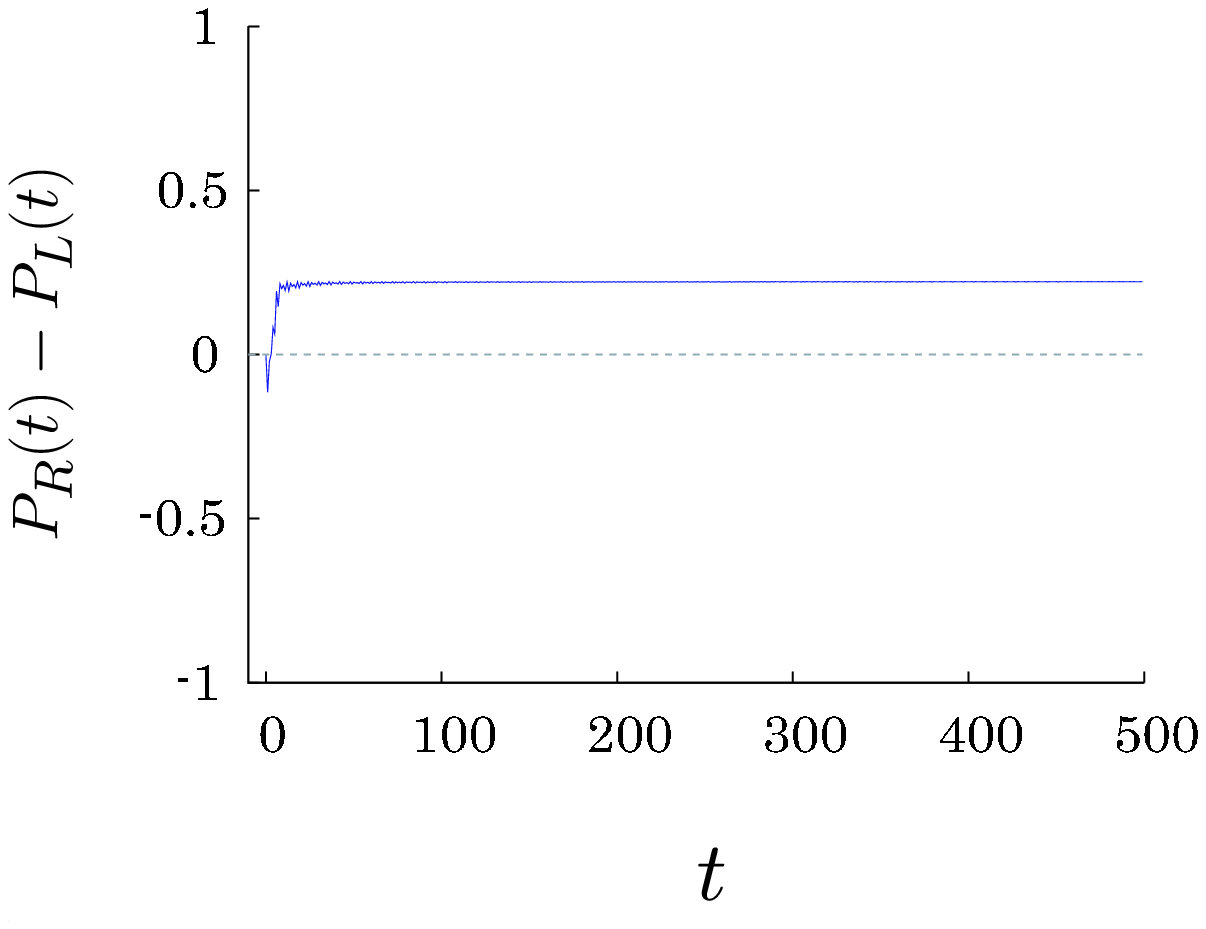}\\[2mm]
  \end{center}
 \end{minipage}
 \bigskip

 (b) $\ket{\Psi_t}=(SXSX)^t\Bigl\{\ket{0}\otimes 1/\sqrt{3}\,\Bigl(\ket{01}+i\ket{10}+\ket{11}\Bigr)\Bigr\}$ with $\theta_1=\pi/6, \theta_2=\pi/4$
 \vspace{5mm}
 \caption{(Blue lines) : Given the initial state $q_0=0, q_1=1/\sqrt{3}, q_2=i/\sqrt{3}, q_3=1/\sqrt{3}$ and the same values of the parameters $\theta_1=\pi/6, \theta_2=\pi/4$, our 2-period time-dependent walk turns to be a losing game in Fig.~(a) and the time-independent walk a winning game in Fig.~(b).}
 \label{fig:fig16}
\end{center}
\end{figure}

\begin{figure}[h]
\begin{center}
 \begin{minipage}{50mm}
  \begin{center}
  \includegraphics[scale=0.4]{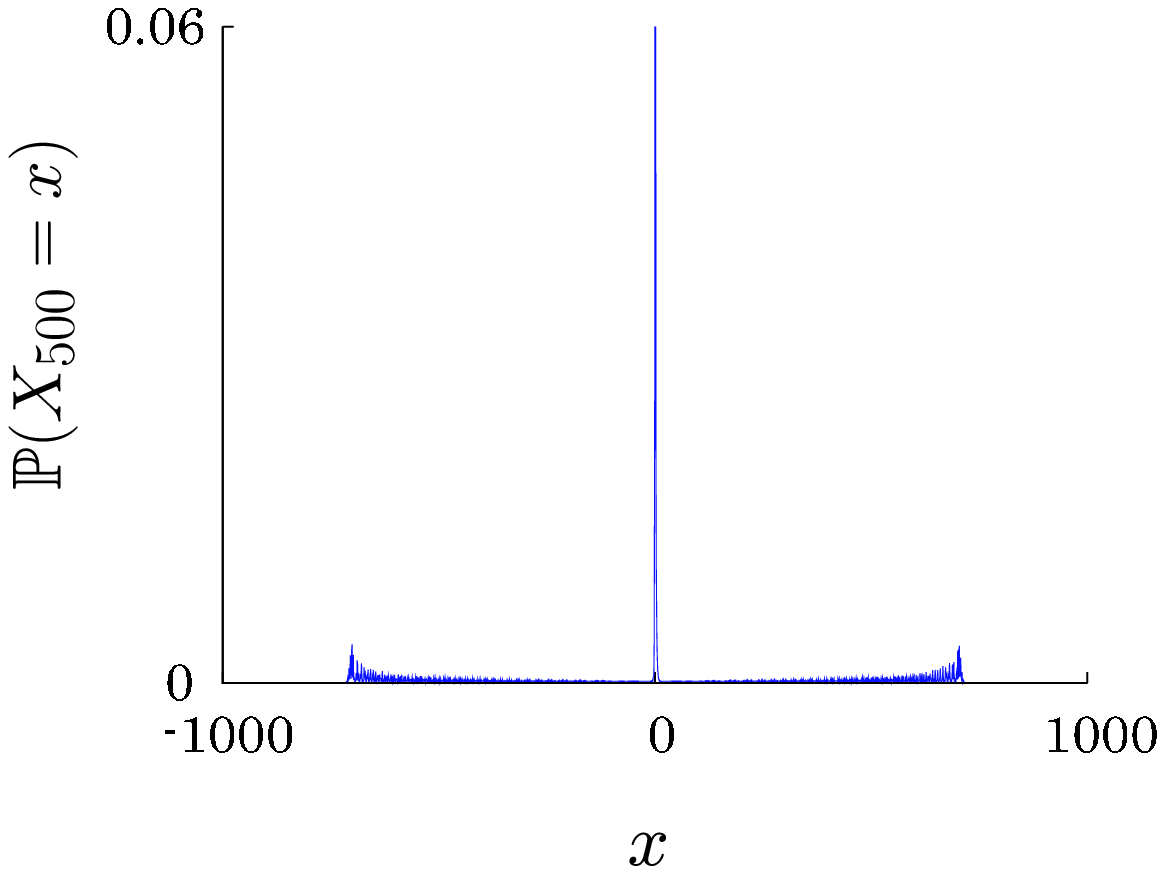}\\[2mm]
  \end{center}
 \end{minipage}
 \begin{minipage}{50mm}
  \begin{center}
 \includegraphics[scale=0.4]{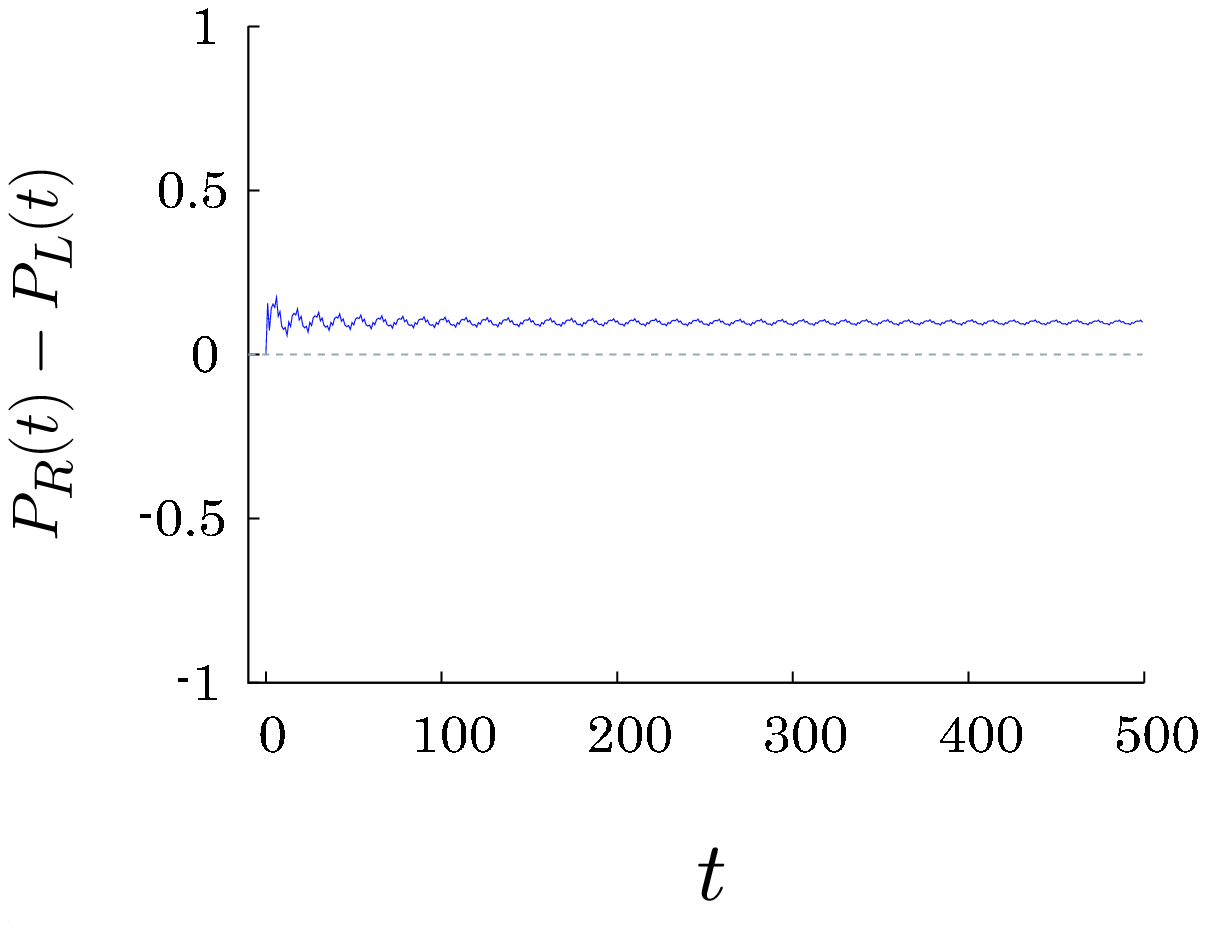}\\[2mm]
  \end{center}
 \end{minipage}
  \bigskip

 (a) $\ket{\Psi_t}=(SYSX)^t(\ket{0}\otimes\ket{01})$ with $\theta_1=\pi/6, \theta_2=\pi/4$
 \vspace{5mm}

 \begin{minipage}{50mm}
  \begin{center}
  \includegraphics[scale=0.4]{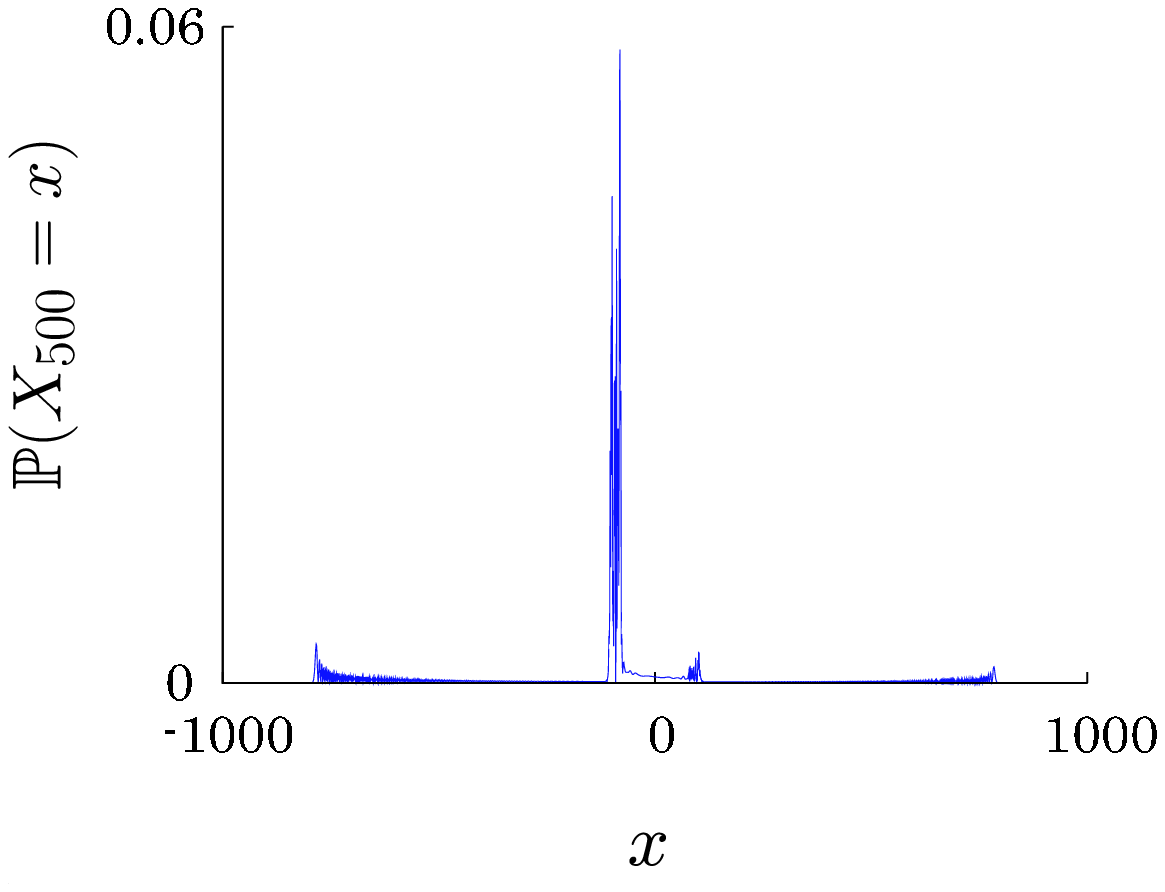}\\[2mm]
  \end{center}
 \end{minipage}
 \begin{minipage}{50mm}
  \begin{center}
 \includegraphics[scale=0.4]{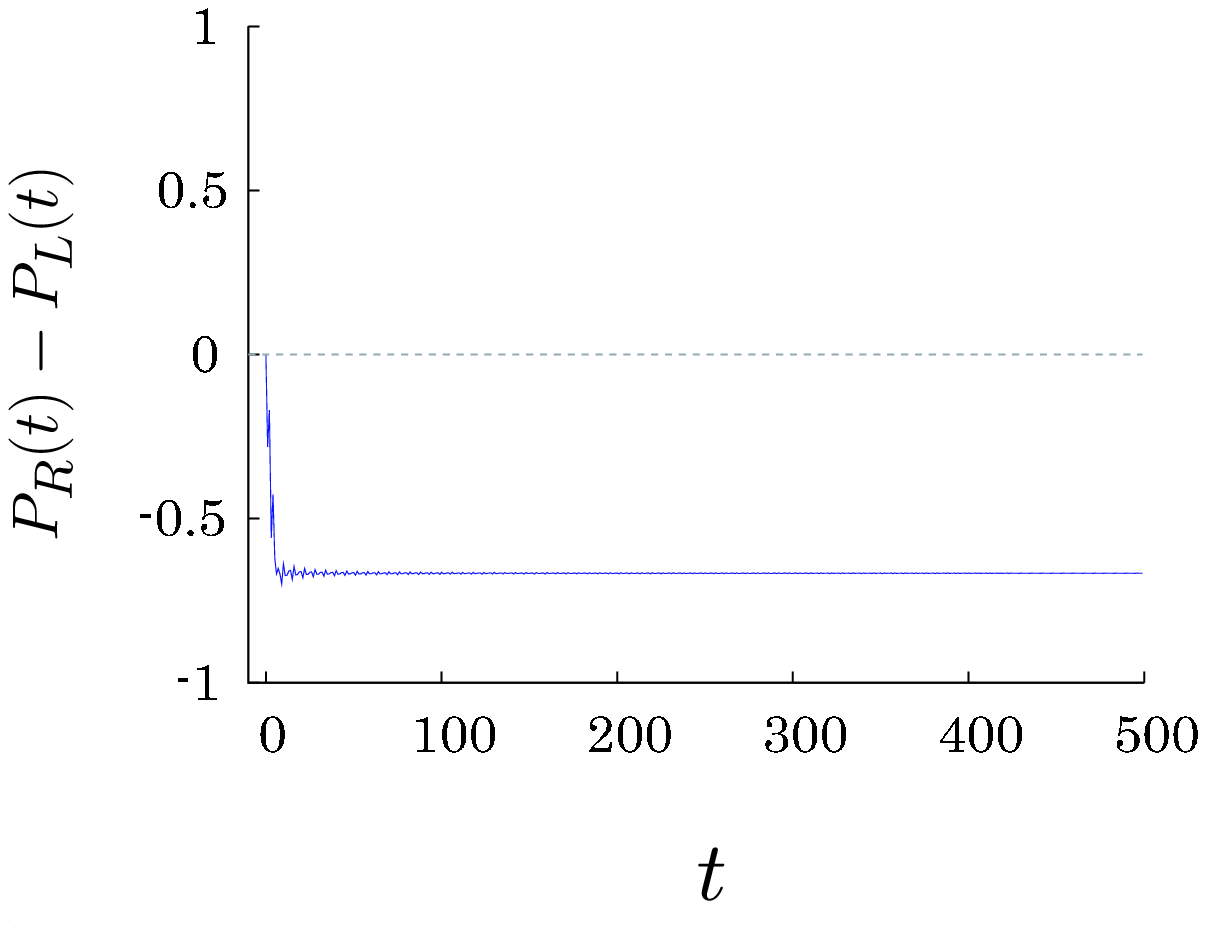}\\[2mm]
  \end{center}
 \end{minipage}
 \bigskip

 (b) $\ket{\Psi_t}=(SYSY)^t(\ket{0}\otimes\ket{01})$ with $\theta_1=\pi/6, \theta_2=\pi/4$
 \vspace{5mm}
 \caption{(Blue lines) : Given the initial state $q_0=0, q_1=1, q_2=0, q_3=0$ and the same values of the parameters $\theta_1=\pi/6, \theta_2=\pi/4$, our 2-period time-dependent walk turns to be a winning game in Fig.~(a) and the time-independent walk a losing game in Fig.~(b).}
 \label{fig:fig17}
\end{center}
\end{figure}

We now consider the game obtained by iterating either
\begin{equation}
 SX\quad \mbox{or}\quad SY,
\end{equation}
which results in the sequence, $U_2 U_2 U_2 U_2\cdots$ for one coin and $U_1 U_1 U_1 U_1\cdots$ for the other one (see Fig.~\ref{fig:fig16}), or $U_1 U_1 U_1 U_1\cdots$ and $U_2 U_2 U_2 U_2\cdots$ (see Fig.~\ref{fig:fig17}).
At this point we can only study this second game by means of numerical simulations.
Our games are very similar to games studied numerically in~\cite{RajendranBenjamin2018}. 
Their results are reported in Fig. 3--a) and Fig. 3--b) of~\cite{RajendranBenjamin2018}.
Figures~\ref{fig:fig16} and \ref{fig:fig17} below give numerical comparisons of our two games and indicate that a Parrondo type paradox is present in our case too.

\section{Summary}
In this paper we analyze a 2-period time-dependent quantum walk on the integers with four internal degrees of freedom.
The walk displays localization and a ballistic spread.
We give two limit theorems for $t\to\infty$.
One of the theorems gives a limit measure which is related to the probability $\mathbb{P}(X_t=x)$, as shown in Figs.~\ref{fig:fig2} and \ref{fig:fig3}.
The other one looks at the convergence of $X_t/t$ in distribution and this is displayed in Fig.~\ref{fig:fig4}.
Our theorems are useful to study aspects of a Parrondo type paradox which was numerically discovered in a quantum game introduced by Rajendran and Benjamin~\cite{RajendranBenjamin2018}. 
A full analytical discussion of this paradox will be pursued in a future publication.
\bigskip\bigskip

\begin{center}
{\bf Acknowledgements}
\end{center}
The second author is supported by JSPS Grant-in-Aid for Young Scientists (B) (No.16K17648).


\end{document}